\documentclass[12pt,letter]{amsart}

\usepackage{amsmath}
\usepackage{amscd}
\usepackage{amssymb}
\usepackage[all]{xy}

\newtheorem{theorem}{Theorem}[section]
\newtheorem{theorem/definition}{Theorem/Definition}[section]

\newtheorem{lemma}{Lemma}[section]

\theoremstyle{remark}

\theoremstyle{definition}

\begin{document}
    \title{Computing Hodge Integrals with one $\lambda$-class}
        \author{Yon-Seo Kim}
        \dedicatory{Department of Mathematics, UCLA}
        \address{University of California, Los Angeles, CA 90095-1555}
        \email{yskim@math.ucla.edu}
        \begin{abstract}
            All Hodge integrals with at-most one $\lambda$-class can be expressed as polynomials in terms of
            lower dimensional Hodge integrals with at-most one $\lambda$-class. Algorithm to compute any
            given Hodge integral with at-most one $\lambda$-class is discussed and some examples are
            presented.
        \end{abstract}
    \maketitle
    \section{Introduction}
    Let $\overline{\mathcal{M}}_{g,n}$ denote the Deligne-Mumford moduli stack of stable curves of genus $g$
    with $n$ marked points. A Hodge integral is an integral of the form
    $$\int_{\overline{\mathcal{M}}_{g,n}}\psi_{1}^{j_{1}}\cdots\psi_{n}^{j_{n}}\lambda_{1}^{k_{1}}
    \cdots\lambda_{g}^{k_{g}}$$
    where $\psi_{i}$ is the first Chern class of the cotangent line bundle at the $i$-th marked point,
    and $\lambda_{1},\cdots,\lambda_{g}$ are the Chern classes of the Hodge bundle. Hodge integrals arise naturally in the
    calculations of Gromov-Witten invariants by localization techniques.
    Their explicit evaluations are difficult problems. The famous \it{Witten's conjecture/Kontsevich's theorem }\rm \cite{Wit},\cite{Kon}
    gives a recursive relation of Hodge integrals involving $\psi$ classes only;
    \begin{equation}\label{eqn:psi}
        \int_{\overline{\mathcal{M}}_{g,n}}\psi_{1}^{j_{1}}\cdots\psi_{n}^{j_{n}}
    \end{equation}
    and some of them can be computed recursively through \it{String Equation }\rm and \it{KdV hierarchy}\rm.
    In \cite{Faber}, C.Faber developed an algorithm to compute intersection numbers of type (\ref{eqn:psi}).
    Also, E.Getzler obtained recursion relations \cite{Getzler} for the case of $g=2$, one of which is given as;
    \begin{align*}
        \langle\langle\tau_{k+2}(x)\rangle\rangle_{2}=&
        \langle\langle\tau_{k+1}(x)\gamma_{a}\rangle\rangle_{0}\langle\langle\gamma^{a}\rangle\rangle_{2}+
        \langle\langle\tau_{k}(x)\gamma_{a}\rangle\rangle_{0}\langle\langle\tau_{1}(\gamma^{a})\rangle\rangle_{2}\\
        &-\langle\langle\tau_{k}(x)\gamma_{a}\rangle\rangle_{0}\langle\langle\gamma^{a}\gamma_{b}\rangle\rangle_{0}\langle\langle\gamma^{b}\rangle\rangle_{2}
        +\frac{7}{10}\langle\langle\tau_{k}(x)\gamma_{a}\gamma_{b}\rangle\rangle_{0}\langle\langle\gamma^{a}\rangle\rangle_{1}\langle\langle\gamma^{b}\rangle\rangle_{1}\\
        &+\frac{1}{10}\langle\langle\tau_{k}(x)\gamma_{a}\gamma_{b}\rangle\rangle_{0}\langle\langle\gamma^{a}\gamma^{b}\rangle\rangle_{1}
        -\frac{1}{240}\langle\langle\tau_{k}(x)\gamma_{a}\rangle\rangle_{1}\langle\langle\gamma^{a}\gamma_{b}\gamma^{b}\rangle\rangle_{0}\\
        &+\frac{13}{240}\langle\langle\tau_{k}(x)\gamma_{a}\gamma^{a}\gamma_{b}\rangle\rangle_{0}\langle\langle\gamma^{b}\rangle\rangle_{1}
        +\frac{1}{960}\langle\langle\tau_{k}(x)\gamma_{a}\gamma^{a}\gamma_{b}\gamma^{b}\rangle\rangle_{0}
        \qquad\text{for }k\geq 0
    \end{align*}
    When the $\lambda$-classes are involved, the computation of Hodge integrals is not easy.
    There was $\lambda_{g}$-conjecture which computes for the case of one top-degree $\lambda$-class as \cite{Fab-Pan};
    \begin{equation}\label{eqn:lambda_g}
        \int_{\overline{\mathcal{M}}_{g,n}}\psi_{1}^{k_{1}}\cdots\psi_{n}^{k_{n}}\lambda_{g}={{2g+n-3}
        \choose{k_{1},\cdots,k_{n}}}\frac{2^{2g-1}-1}{2^{2g-1}}\frac{\vert B_{2g}\vert}{(2g)!}
    \end{equation}
    where $B_{2g}$ are Bernoulli numbers and $k_{1}+\cdots+k_{n}=2g-3+n$.
    By using Mari\~{n}o-Vafa formula \cite{Mar-Vaf}\cite{LLZ1}, the case of $\lambda_{g-1}$ with one marked point can be computed as;
    \begin{equation}\label{eqn:g-1}
        \int_{\overline{\mathcal{M}}_{g,1}}\psi_{1}^{2g-1}\lambda_{g-1}=b_{g}\sum_{i=1}^{2g-1}\frac{1}{i}-\frac{1}{2}
        \sum_{g_{1}+g_{2}=g}\frac{(2g_{1}-1)!(2g_{2}-1)!}{(2g-1)!}b_{g_{1}}b_{g_{2}}
    \end{equation}
    and the case of more than one marked points can be computed by repeatedly applying the Cut-and-Join equation:
    \begin{equation}\label{eqn:CJ}
        \frac{\partial\Omega}{\partial\tau}=\frac{\sqrt{-1}\lambda}{2}\sum_{i,j\geq 1}\Big(ijp_{i+j}
        \frac{\partial^{2}\Omega}{\partial p_{i}\partial p_{j}}+ijp_{i+j}\frac{\partial\Omega}{\partial p_{i}
        }\frac{\partial\Omega}{\partial p_{j}}+(i+j)p_{i}p_{j}\frac{\partial\Omega}{\partial p_{i+j}}\Big)
    \end{equation}
    which was used in the proof of Mari\~{n}o-Vafa formula, and proven to be an effective tool in studying Hodge integrals.

    The moduli space of relative stable morphisms admits a natural $S^{1}$-action induced from the $S^{1}$-action on the target space.
    And as a result of the localization formula applied to it, the following convolution formula is obtained; \\

    \bf{Theorem 6.1. }\it
        For any partition $\mu$ and $e$ with $\vert e\vert<\vert\mu\vert+l(\mu)-\chi$, we have
        \begin{equation*}
            \big[\lambda^{l(\mu)-\chi}\big]\sum_{\mid\nu\mid=\mid\mu\mid}\Phi^{\bullet}_{\mu,\nu}(-\lambda)
            z_{\nu}\mathcal{D}_{\nu,e}^{\bullet}(\lambda)=0
        \end{equation*}
        where the sum is taken over all partitions $\nu$ of the same size as $\mu$.
    \rm\\

    Here $\chi$ is the prescribed Euler number of domain curves,
    $\big[\lambda^{a}\big]$ means taking the coefficient of $\lambda^{a}$,
    $\Phi^{\bullet}(\lambda)$ is a generating series of Double Hurwitz Numbers, and
    $\mathcal{D}^{\bullet}(\lambda)$ is a certain generating series of Hodge integrals.
    This formula gives many relations between Hodge integrals with at-most one $\lambda$-class,
    and it is enough to consider the special case of $\mu=(d)$ for some positive integers $d$
    to compute all Hodge integrals with at-most one $\lambda$-class.
    More precisely, the following theorem is proved in Section 7; \\

    \bf{Theorem 7.2. }\it
            Any given Hodge integral with one $\lambda$-class:
            \begin{equation*}\label{eqn:one_lambda}
                \int_{\overline{\mathcal{M}}_{g,n}}\psi_{1}^{k_{1}}\cdots\psi_{n}^{k_{n}}\lambda_{j}
            \end{equation*}
            where $k_{1},\cdots,k_{n}\in\mathbb{N}\cup\{0\}$, $j\in\{0,1,2,\cdots,g\}$,
            is explicitly expressed as a polynomial in terms of lower-dimensional Hodge integrals with one $\lambda$-class.
            Therefore it computes all Hodge integrals with one $\lambda$-class.
    \rm\\

    The rest of the paper is organized as follows:
    In Section 2, we summarize various versions of localization formulas which will be used in computing
    Hodge integrals. In Section 3, the Relative Moduli Space is defined and the natural $S^{1}$-action on it is introduced.
    Also the fixed locus of the $S^{1}$-action and their corresponding description in terms of graphs is discussed.
    In Section 4, we compute the Euler class of the normal bundle of the fixed locus of the $S^{1}$-action
    in the relative moduli space. In Section 5, the Double Hurwitz Numbers and its description in terms of
    Hodge integrals over a certain moduli space is discussed.
    In Section 6, the Recursion Formula which gives relations between Hodge integrals with at-most one $\lambda$-class is proved.
    In Section 7, the Recursion Formula is used prove that all Hodge integrals with at-most one $\lambda$-class is explicitly expressed
    as a polynomial in terms of lower dimensional Hodge integrals with at-most one $\lambda$-class.
    In Section 8, an algorithm to implement the Recursion Formula and to compute each Hodge integral
    with at-most one $\lambda$-class is discussed.
    In Section 9, some examples of the algorithm in section 8 are presented.\\

    {\bf Acknowledgements. } {\em
        I would like to thank my advisor, Professor Kefeng Liu for his constant support and
        encouragements as well as many inspiring discussions.
        I would also like to thank Xiaowei Wang for helpful discussions and friendship.}
%----------------------------------------------------------------------------------------------
    \vskip 10mm
    \section{Localization Formula}
        In this section, I will summarize various versions of localization formulas.
        \subsection{Equivariant Cohomology}
            Let $G$ be a compact Lie group acting on $M$. The equivariant cohomology of $M$ is defined as the
            ordinary cohomology of the space $M_{G}$ obtained from a fixed universal $G$-bundle $EG$, by the
            mixing construction $$M_{G}=EG\times_{G}M$$ Here, $G$ acts on the right of $EG$ and on the left
            of $M$, and the notation means that we identify $(pg,q)\thicksim(p,gq)$ for $p\in EG$, $q\in M$,
            $g\in G$. Hence $M_{G}$ is the bundle with fibre $M$ over the classifying space $BG$ associated
            to the universal bundle $EG\longrightarrow BG$. We have natural projection map $\pi:M_{G}
            \longrightarrow BG$ and $\sigma:M_{G}\longrightarrow M/G$, which fits into the mixing diagram
            of Cartan and Borel:
            \begin{equation*}
                \xymatrix{EG\ar[d] & EG\times M\ar[r]\ar[d]\ar[l] & M\ar[d] \\ BG & E\times_{G}M\ar[l]_{\pi}
                \ar[r]^{\sigma} & M/G \\}
            \end{equation*}
            If $G$ acts smoothly on $M$, then we have $M_{G}\cong M/G$. This is not true in general but it
            turns out that $M_{G}$ is a better functorial construction and the proper homotopy theoretic
            quotient of $M$ by $G$. In any case, the equivariant cohomology, denoted by $H_{G}^{*}(M)$,
            is defined by $$H_{G}^{*}(M)=H^{*}(M_{G})$$ and constitutes a contravariant functor from
            $G$-spaces to modules over the base ring $H_{G}^{*}:=H_{G}^{*}(pt)=H^{*}(BG)$. The map $\sigma$
            defines a natural map $\sigma^{*}:H^{*}(M/G)\longrightarrow H_{G}^{*}(M)$ which is an
            isomorphism if $G$ acts freely. The inclusion $i:M\longrightarrow M_{G}$ induces a natural map
            $i^{*}:H_{G}^{*}(M)\longrightarrow H^{*}(M)$.
        \subsection{Atiyah-Bott Localization Formula}
            Let $i:V\hookrightarrow M$ be a map of compact manifolds. The tubular neighborhood of $V$ inside
            $M$ can be identified with the normal bundle of $V$. On the total space of the normal bundle,
            there is the Thom form $\Phi_{V}$ which has compact support in the fibres and integrates to one
            in each fiber. Extending this form by zero gives a form in $M$, and multiplying by $\Phi_{V}$
            provides a map $H^{*}(V)\cong H^{*+k}(M,M\backslash V)\longrightarrow H^{*}(M)$. In particular,
            the cohomology class $1\in H^{0}(V)$ is sent to the Thom class and this class restricts to be
            the Euler class of the normal bundle of $V$ in $M$, $\mathcal{N}_{V/M}$. Hence, we see that
            $$i^{*}i_{*}1=e(\mathcal{N}_{V/M})$$
            This also holds in equivariant cohomology by same argument applied to $V_{G},M_{G}$.
            The theorem of Atiyah and Bott says that an inverse of the Euler class of the normal bundle
            always exists along the fixed locus of a group action. Precisely, $i^{*}/e(\mathcal{N}_{V/M})$
            is the inverse of $i_{*}$ in equivariant cohomology, i.e. for any equivariant class $\phi$,
            $$\phi=\sum_{F}\frac{i_{*}i^{*}\phi}{e_(\mathcal{N}_{F/M})}$$
            holds where $F$ runs over the fixed locus of the group action. In the integrated form, we have
            $$\int_{M}\phi=\sum_{F}\int_{F}\frac{i^{*}\phi}{e(\mathcal{N}_{F/M})}$$ \\

        \subsection{Functorial Localization Formula}
            Let $X$ and $Y$ be $T$-manifolds. Assume $f:X\longrightarrow Y$ is a $T$-equivariant map,
            $j_{E}:E\hookrightarrow Y$ is a fixed component in $Y$, and $i_{F}:F\hookrightarrow f^{-1}(E)$ is a fixed component in $X$.
            For any equivariant class $\omega\in H_{T}^{*}(X)$, we have the diagrams;
            \begin{align*}
                \xymatrix@R.6in{ F\ar[rr]^{i_{F}}\ar[d]^{g=f\mid_{F}} & &X\ar[d]^{f}\\E\ar[rr]^{j_{E}} & & Y \\}
                &&
                \xymatrix{\frac{i_{F}^{*}(\omega)}{e_{T}(F/X)}\ar[d]^{g_{!}} & &\omega\ar[ll]_{i_{F}^{*}}
                \ar[d]^{f_{!}} \\ g_{!}\Big[\frac{i_{F}^{*}(\omega)}{e_{T}(F/X)}\Big] & &
                f_{!}(\omega)\ar[ll]_{j_{E}^{*}}}
            \end{align*}
            Applying the Atiyah-Bott Localizatio Formula with the naturality relation $f_{!}(\omega\cdot f^{*}\alpha)=f_{!}\omega\cdot\alpha$,
            we obtain the Functorial Localization Formula:
            \begin{equation}\label{formula:func_loc}
                g_{!}\Big[\frac{i_{F}^{*}(\omega)}{e_{T}(F/X)}\Big]=\frac{j_{E}^{*}f_{!}(\omega)}{e_{T}(E/Y)}
            \end{equation}\\

        \subsection{Virtual Functorial Localization Formula}
            The above Functorial Localization Formula is also valid in the case where $X$ and $F$ are virtual
            fundamental classes. In this paper, we will use $\big[\overline{\mathcal{M}}_{\chi,n}^{\bullet}(\mathbb{P}^
            {1},\mu)\big]^{vir}$ for $X$, and $\big[F_{\Gamma}\big]^{vir}$ for $F$. Hence for any equivariant
            class $\omega$, we have:
            \begin{equation}\label{formula:vir_func_loc}
                \int_{\big[\overline{\mathcal{M}}_{\chi,n}^{\bullet}(\mathbb{P}^{1},\mu)\big]^{vir}}\omega
                =\sum_{F_{\Gamma}}\int_{\big[F_{\Gamma}\big]^{vir}}\frac{i_{\Gamma}^{*}(\omega)}{e_{T}(F_{\Gamma}
                /\overline{\mathcal{M}}^{\bullet}_{\chi}(\mathbb{P}^{1},\mu))}
            \end{equation} \\
%-----------------------------------------------------------------------------------------
    \vskip 10mm
    \section{Relative Moduli Space and $S^{1}$-action}
        \subsection{Moduli space of relative morphisms}
            For any non-negative integer $m$, let $$\mathbb{P}^{1}[m]=\mathbb{P}^{1}_{(0)}\cup
            \mathbb{P}^{1}_{(1)}\cup\cdots\cup\mathbb{P}^{1}_{(m)}$$ be a chain of $m+1$ copies
            of $\mathbb{P}^{1}$ such that $\mathbb{P}^{1}_{(l)}$ is glued to $\mathbb{P}^{1}_{(l+1)}$
            at $p_{1}^{(l)}$ for $0\leq l\leq m-1$.
            The irreducible component $\mathbb{P}^{1}_{(0)}$ is referred to as \it{the root component}\rm,
            and other irreducible components are called \it{bubble components}\rm. Two points $p_{1}^{(l)}
            \neq p_{1}^{(l+1)}$ in $\mathbb{P}^{1}_{(l)}$ are fixed.
            Denote by $\pi[m]:\mathbb{P}^{1}[m]\longrightarrow\mathbb{P}^{1}$ the map which is identity
            on the root component and contracts all bubble components to $\mathbb{P}_{1}^{(0)}$.
            Also denote by $\mathbb{P}^{1}(m)=\mathbb{P}^{1}_{(1)}\cup\cdots\cup\mathbb{P}^{1}_{(m)}$
            the union of bubble components of $\mathbb{P}^{1}[m]$. \\
            For a fixed partition $\mu$ of a positive integer $d$,
            let $\overline{\mathcal{M}}_{\chi,n}^{\bullet}(\mathbb{P}^{1},\mu)$ be \it{the moduli space of relative
            morphisms  }\rm$f:\big( C; x_{1},\cdots,x_{l(\mu)},z_{1},\cdots,z_{n}\big)\longrightarrow\big(\mathbb{P}
            ^{1}[m],p_{1}^{(m)}\big)$ such that
            \begin{itemize}
                \item[(1)] $\big(C;x_{1},\cdots,x_{l(\mu)},z_{1},\cdots,z_{n}\big)$ is a possibly-disconnected
                        prestable curve of Euler number $\chi$ with $l(\mu)+n$ marked points.
                        Here, the marked points are \it{unordered}\rm.
                \item[(2)] $f^{-1}(p_{1}^{(m)})=\sum_{i=1}^{l(\mu)}\mu_{i}x_{i}$ as Cartier divisors
                        and $\text{deg}(\pi[m]\circ f)=\vert\mu\vert$.
                \item[(3)] The preimage of each node in $\mathbb{P}^{1}[m]$ consists of nodes of $C$.
                        If $f(y)=p_{1}^{(l)}$ and $C_{1}$ and $C_{2}$ are two irreducible components of
                        $C$ which intersects at $y$, then $f\mid_{C_{1}}$ and $f\mid_{C_{2}}$ have the
                        same contact order to $p_{1}^{(l)}$ at $y$.
                \item[(4)] The automorphism group of $f$ is finite. Here, an automorphism of $f$ consists
                        of an automorphism of the domain curve and and automorphism of the pointed curve
                        $\big(\mathbb{P}^{1}(m),p_{1}^{(0)},p_{1}^{(m)}\big)$.
            \end{itemize}
            Following \cite{Li1}\cite{Li2}, $\overline{\mathcal{M}}_{\chi,n}^{\bullet}(\mathbb{P}^{1}
            ,\mu)$ is a separated, proper Deligne-Mumford stack with a perfect obstruction theory of virtual
            dimension $r=-\chi+\vert\mu\vert+l(\mu)+n$, and hence has a virtual fundamental class of degree
            $r$.
        \subsection{Torus Action}
            Consider the $\mathbb{C}^{*}$-action $t\cdot[z^{0}:z^{1}]=[tz^{0}:z^{1}]$ on $\mathbb{P}^{1}$.
            There are two fixed points $p_{0}=[0:1]$ and $p_{1}=[1:0]$. Extend this action to the action on
            $\mathbb{P}^{1}[m]$ by identifying the root component with $\mathbb{P}^{1}$ and giving trivial
            actions on bubble components. Then this extended action on $\mathbb{P}^{1}[m]$ induces
            an action on $\overline{\mathcal{M}}_{\chi,n}^{\bullet}(\mathbb{P}^{1},\mu)$.
        \subsection{Fixed Locus}
            The connected components of the fixed points set of $\overline{\mathcal{M}}_{\chi,n}^{\bullet}
            (\mathbb{P}^{1},\mu)$ under the induced torus action can be parametrized by labeled graphs.
            For any $f:\big( C; x_{1},\cdots,x_{l(\mu)},z_{1},\cdots,z_{n}\big)\longrightarrow\mathbb{P}^{1}[m]$
            representing a fixed point of the $\mathbb{C}^{*}$-action on $\overline{\mathcal{M}}_{\chi,n}^{\bullet}(
            \mathbb{P}^{1},\mu)$, the restriction of $\hat{f}:=\pi[m]\circ f:C\longrightarrow\mathbb{P}^{1}
            $ to an irreducible component of $C$ is either a constant map to one of the $\mathbb{C}^{*}$
            -fixed points $p_{0},p_{1}$, or a covering of $\mathbb{P}^{1}$ fully ramified over $p_{0}$ and
            $p_{1}$. Associate a labeled graph $\Gamma$ to the $\mathbb{C}^{*}$-fixed point
            $\big[\,\, f:\big( C;x_{1},\cdots,x_{l(\mu)},z_{1},\cdots,z_{n}\big)\longrightarrow\mathbb{P}^{1}[m]\,\,\big]$
            as follows:
            \begin{itemize}
                \item[(1)] For each connected component $C_{v}$ of $\hat{f}^{-1}(\{p_{0},p_{1}\})$,
                    assign a \it{vertex }\rm $v$, a label $g(v)$ which is the arithmetic genus of $C_{v}$,
                    , and a label $i(v)=\begin{cases} 0,&\text{if }\hat{f}(C_{v})=p_{0} \\ 1,&\text{if }
                    \hat{f}(C_{v})=p_{1}\end{cases}$.
                    Denote by $V(\Gamma)^{(k)}$ the set of vertices $v$
                    with $i(v)=k$, for $k=0,1$. The set $V(\Gamma)$ of vertices of the graph $\Gamma$ will
                    then be the disjoint union of $V(\Gamma)^{(0)}$ and $V(\Gamma)^{(1)}$.
                \item[(2)] Assign an \it{edge }\rm $e$ to each rational irreducible component $C_{e}$ of $C$
                    such that $\hat{f}\mid_{C_{e}}$ is not a constant map. Then $\hat{f}\mid_{C_{e}}$
                    is fully ramified over $p_{0}$ and $p_{1}$ with degree $d(e)$. Let $E(\Gamma)$ denote
                    the set of edges of $\Gamma$.
                \item[(3)] The set of flags is given by
                    $F(\Gamma)=\{(v,e):v\in V(\Gamma),e\in E(\Gamma), C_{v}\cap C_{e}\neq\emptyset\}$
                \item[(4)] For each $v\in V(\Gamma)$, define $d(v)=\sum_{(v,e)\in F(\Gamma)}d(e)$
                    and let $\nu(v)$ be the partition of $d(v)$ determined by $\{d(e):(v,e)\in F(\Gamma)\}$.
                    In case $m>0$, we assign an additional label for each $v\in V(\Gamma)^{(1)}$: let
                    $\mu(v)$ be the partition of $d(v)$ determined by the ramification of $f\mid_{C_{v}}:
                    C_{v}\longrightarrow\mathbb{P}^{1}(m)$ over $p_{1}^{(m)}$.
            \end{itemize}
            Let $G_{\chi}(\mathbb{P}^{1},\mu)$ be the set of all the graphs associated to the
            $\mathbb{C}^{*}$-fixed points in $\overline{\mathcal{M}}_{\chi,n}^{\bullet}(\mathbb{P}^{1},\mu)$.
            We now describe the set of fixed points associated to a given graph
            $\Gamma\in G_{\chi}(\mathbb{P}^{1},\mu)$.
            \begin{itemize}
                \item{\underline{Case $m=0$:}} Any $\mathbb{C}^{*}$-fixed point in $\overline{\mathcal{M}}
                    _{\chi,n}^{\bullet}(\mathbb{P}^{1},\mu)$ which is represented by a morphism to
                    $\mathbb{P}^{1}[0]=\mathbb{P}^{1}$ is associated to a graph $\Gamma_{0}\in G_{\chi}^{0}
                    (\mathbb{P}^{1},\mu)$ such that
                    \begin{align*}
                        &V(\Gamma_{0})^{(0)}=\{v^{0}_{1},\cdots,v^{0}_{k}\},\quad g(v^{0}_{i})=g_{i},
                            \quad k\in\mathbb{N},\qquad\sum_{i}(2-2g_{i})=\chi \\
                        &V(\Gamma_{0})^{(1)}=\{ v^{\infty}_{1},\cdots,v^{\infty}_{l(\mu)}\},\quad
                            g(v^{\infty}_{1})=\cdots=g(v^{\infty}_{l(\mu)})=0,\\
                        &E(\Gamma_{0})=\{e_{1},\cdots,e_{l(\mu)}\},\quad d(e_{i})=\mu_{i}
                            \text{ for }i=1,\cdots,l(\mu)\\
                        &j(v)=\text{number of }z_{i}\text{'s mapped to }v,\quad \sum_{V(\Gamma_{0})^{(0)}}j(v)=n
                    \end{align*}
                    The two end-points of the edge $e_{i}$ are $v^{0}_{j}$ and $v^{\infty}_{i}$ for some 
                    $1\leq j\leq k$. Let $\mu(v_{i}^{0})=\{\mu_{j}\vert e_{j}\text{ has }v_{i}^{0}
                    \text{ as an endpoint }\}$. Define
                    $$\overline{\mathcal{M}}_{\Gamma_{0}}=\prod_{1\leq i\leq k}\overline{\mathcal{M}}_{g_{i},
                    l(\mu(v_{i}^{0}))+j(v_{i}^{0})}$$ where we take $\overline{\mathcal{M}}_{0,1}=\overline{\mathcal{M}}
                    _{0,2}=\overline{\mathcal{M}}_{1,0}=\{\text{pt}\}$, then there is a morphism $i_{\Gamma_
                    {0}}:\overline{\mathcal{M}}_{\Gamma_{0}}\longrightarrow\overline{\mathcal{M}}_{\chi,n}
                    ^{\bullet}(\mathbb{P}^{1},\mu)$ whose image is the fixed locus $F_{\Gamma_{0}}$
                    associated to $\Gamma_{0}$. Hence $i_{\Gamma_{0}}$ induces an isomorphism $\overline{
                    \mathcal{M}}_{\Gamma_{0}}/A_{\Gamma_{0}}\cong F_{\Gamma_{0}}$ where $A_{\Gamma_{0}}$ is
                    the automorphism group of any morphism associated to the graph $\Gamma_{0}$, which can
                    be obtained from the short exact sequence 
                    $$1\longrightarrow\prod^{l(\mu)}_{i=1}\mathbb{Z}_{\mu_{i}}\longrightarrow A_{\Gamma^{0}}
                    \longrightarrow\text{Aut }(\Gamma_{0})\longrightarrow 1$$
                    The virtual dimension of $F_{\Gamma_{0}}$ with only stable vertices is
                    $$d_{\Gamma_{0}}=-\frac{3}{2}\chi+l(\mu)+n$$
                    For any vertex $v\in V(\Gamma_{0})^{(0)}$, introduce \it{multiplicity of $v$}\rm, $m(v)$, as follows:
                    $$m(v)=\vert\{w\in V(\Gamma_{0})^{(0)}\,\mid\,\,g(v)=g(w),\,\mu(v)=\mu(w),\,j(v)=j(w)\}\vert$$
                    Pick one representatives from each group of identical vertices $\{v_{1},\cdots,v_{l}\}$ so that
                    $\sum m(v_{k})=\vert V(\Gamma_{0})^{(0)}\vert$. Denote by $m(v_{k})=m_{k}$ and now we can find
                    the order of automorphism group of any $\Gamma_{0}\in G_{\chi,n}^{0}(\mathbb{P}^{1},\mu)$ to be:
                    \begin{equation}\label{formula:aut_zero}
                        \vert\text{Aut }\Gamma_{0}\vert=\prod_{k}m_{k}!\,\vert\,j(v_{k})!\,\,\text{Aut }\mu(v_{k})\,\vert^{m_{k}}
                    \end{equation}

                \item{\underline{Case $m>0$:}}
                    For any given $f:(C;x_{1},\cdots,x_{l(\mu)},z_{1},\cdots,z_{n})\longrightarrow\mathbb{P}^{1}[m]$,
                    consider $f_{0}$ and $f_{\infty}$ which are defined as follows:
                    \begin{itemize}
                        \item{\underline{$f_{0}$}:} Let $C_{0}=f^{-1}(p^{(0)}_{0})$, $\{y_{1},\cdots,y_{l(\nu)}\}=f^{-1}(p_{1}^{(0)})$,
                            and $\{z_{1},\cdots,z_{n_{0}}\}$ mapped to $C_{0}$.
                            Then $f_{0}:(C_{0};y_{1},\cdots,y_{l(\nu)},z_{1},\cdots,z_{n_{0}})\longrightarrow\mathbb{P}^{1}$ corresponds
                            to the case of $m=0$.
                        \item{\underline{$f_{\infty}$}:} Let $C_{\infty}=f^{-1}(\mathbb{P}^{1}(m))$ and
                            $f_{\infty}=f\mid_{C_{\infty}}$, then $f_{\infty}$ corresponds to an element of
                            $\overline{\mathcal{M}}_{\Gamma}^{(1)}$ corresponding to the graph $\Gamma$ described below.
                    \end{itemize}
                    \vskip 5mm
                    Classify the vertices of the graph $\Gamma$ as follows:
                    \begin{align*}
                        &V^{I}(\Gamma)^{(0)}= \{v\in V(\Gamma)^{(0)}:r_{0}(v)=-1\},\\
                        &V^{II}(\Gamma)^{(i)}= \{v\in V(\Gamma)^{(i)}:r_{i}(v)=0\},&&\text{for }i=0,1 \\
                        &V^{S}(\Gamma)^{(i)}= \{v\in V(\Gamma)^{(i)}:r_{i}(v)>0\},&&\text{for }i=0,1 \\
                        \text{where }\,\,&r_{0}(v)= 2g(v)-2+\text{val}(v)+j(v), &&\text{for }v\in V(\Gamma)^{(0)}, \\
                        &r_{1}(v)= 2g(v)-2+l(\mu(v))+l(\nu(v))+j(v), &&\text{for }v\in V(\Gamma)^{(1)}
                    \end{align*}
                    Define $\overline{\mathcal{M}}_{\Gamma}=\overline{\mathcal{M}}_{\Gamma}^{(0)}\times
                    \overline{\mathcal{M}}_{\Gamma}^{(1)}$, where
                    $\overline{\mathcal{M}}_{\Gamma}^{(0)}=\prod_{v\in V^{S}(\Gamma)^{(0)}}
                    \overline{\mathcal{M}}_{g(v),\text{val}(v)+j(v)}$ and
                    $\overline{\mathcal{M}}_{\Gamma}^{(1)}$ is the moduli space of morphisms
                    $\hat{f}:\hat{C}\longrightarrow\mathbb{P}^{1}(m)$ such that
                    \begin{itemize}
                        \item[(1)] $\hat{C}=\bigsqcup_{v\in V(\Gamma)^{(1)}}C_{v}$
                        \item[(2)] For each $v\in V(\Gamma)^{(1)}$, $\big(C_{v};x_{v,1},\cdots,
                            x_{v,l(\mu(v))},y_{v,1},\cdots,y_{v,l(\nu(v))},z_{1},\cdots,z_{j(v)}\big)$
                            is a prestable curve of genus $g(v)$ with $l(\mu(v))+l(\nu(v))+j(v)$ ordered marked points.
                        \item[(3)] As Cartier divisors, \\
                            $(\hat{f}\mid_{C_{v}})^{-1}(p_{1}^{(0)})=\sum^{l(\nu(v))}_{j=1}\nu(v)_{j}
                            y_{v,j},\quad (\hat{f}\mid_{C_{v}})^{-1}(p_{1}^{(m)})=\sum^{l(\mu(v))}_{i=1}
                            \mu(v)_{i}x_{v,i}$ \\
                            The morphism $(\hat{f}\mid_{C_{v}})^{-1}(E)\longrightarrow E$ is of degree
                            $d(v)$ for each irreducible component $E$ of $\mathbb{P}^{1}(m)$.
                        \item[(4)] The automorphism group of $\hat{f}$ is finite. Here, an automorphism of
                            $\hat{f}$ consists of an automorphism of the domain curve $\hat{C}$ and an
                            automorphism of the pointed curve $(\mathbb{P}^{1}(m),p_{1}^{(0)},p_{1}^{(m)})$
                            , which is an element of $(\mathbb{C}^{*})^{m}$.
                    \end{itemize}
                    There is a morphism $i_{\Gamma}:\overline{\mathcal{M}}_{\Gamma}\longrightarrow
                    \overline{\mathcal{M}}_{\chi,n}^{\bullet}(\mathbb{P}^{1},\mu)$ whose image is the fixed
                    locus $F_{\Gamma}$ associated to the graph $\Gamma$. Hence $i_{\Gamma}$ induces an
                    isomorphism $\overline{\mathcal{M}}_{\Gamma}/A_{\Gamma}\cong F_{\Gamma}$ where
                    $A_{\Gamma}$ is the automorphism group of any morphism associated to the graph $\Gamma$,
                    which can be obtained from the short exact sequence
                    $$1\longrightarrow\prod_{e\in E(\Gamma)}\mathbb{Z}_{d(e)}\longrightarrow A_{\Gamma}
                    \longrightarrow\text{Aut}(\Gamma)\longrightarrow 1$$
                    The virtual dimension of $\overline{\mathcal{M}}^{(1)}_{\Gamma}$ is given by
                    $d_{\Gamma}^{(1)}=\Big(\sum_{v\in V(\Gamma)^{(1)}}r_{1}(v)\Big)-1$. Use the identities
                    $$-\chi=-\chi_{0}-\chi_{\infty}+2l(\nu),\qquad g=%\sum_{v\in V(\Gamma)}g(v)+b_{1}(\Gamma)=
                    \sum_{v\in V(\Gamma)}g(v)-\mid V(\Gamma)\mid+\mid E(\Gamma)\mid +1$$
                    to get the virtual dimension of $F_{\Gamma}$:
                    \begin{align*}
                    d_{\Gamma}=&d_{\Gamma}^{(0)}+d_{\Gamma}^{(1)}=-\frac{3}{2}\chi_{0}+l(\nu)-\chi_{\infty}+l(\mu)+l(\nu)-1+n_{\infty} \\
                        =&\sum_{v\in V^{S}(\Gamma)^{(0)}}\big(3g(v)-3+\text{val}(v)\big)+\big(\sum_{v\in V(\Gamma)^{(1)}}r_{1}(v)\big)-1+n_{\infty} \\
                        =&2g-3+l(\mu)+\sum_{v\in V^{S}(\Gamma)^{(0)}}\big(g(v)-1\big)+\mid V^{I}(\Gamma)^{(0)}\mid+n_{\infty} \\
                        =&-\frac{1}{2}\chi_{0}-\chi+l(\mu)-1+n_{\infty}
                    \end{align*}
                    By similar observation as in the case of $m=0$, we find the order of automorphism group of any given
                    $\Gamma\in G_{\chi}^{\infty}(\mathbb{P}^{1},\mu)$ to be:
                    \begin{equation}\label{formula:aut_infty}
                        \vert\text{Aut }\Gamma\vert=\vert\text{Aut }\Gamma_{0}\vert\,\,
                        \vert\text{Aut }\mu\,\vert\Big(\prod n_{k}!\Big)\Big(\prod_{V(\Gamma)^{(1)}}j(v)!
						\,\,\vert\text{Aut }\mu(v)\vert\,\,\vert\text{Aut }\nu(v)\vert\Big)
                    \end{equation}
                    In this case, $n_{k}$ is the multiplicity of vertices in $V(\Gamma)^{(1)}$ with identical
                    $\mu(v)$, $\nu(v)$, $j(v)$, and $g(v)$.
                    The presence of automorphism group of $\mu$ is due to the fact
                    that we can exchange two marked points with same ramification type $\mu_{i}$ without changing
                    the type of corresponding graph.
            \end{itemize}
%----------------------------------------------------------------------------------------------
    \vskip 10mm
    \section{Computation of $e_{T}(\mathcal{N}_{\Gamma}^{vir})$}
        In this section, I will summarize the computations of $e_{T}(\mathcal{N}_{\Gamma}^{vir})$
        in \cite{LLZ1} which will be needed for localization computation.
        Denote by $(\omega)$ the 1-dimensional representation of $\mathbb{C}^{*}$ given by
        $\lambda\cdot z=\lambda^{\omega}z$ for $\lambda\in\mathbb{C}^{*}$, $z\in\mathbb{C}$.
        For a given graph $\Gamma\in G_{g,n}(\mathbb{P}^{1},\mu)$, let
        \begin{equation}\label{fixed_point}
            \big[\,\,f:(C,x_{1},\cdots,x_{l(\mu)},z_{1},\cdots,z_{n})\longrightarrow\mathbb{P}^{1}[m]\,\,\big]
        \end{equation}
        be a fixed point of the $\mathbb{C}^{*}$-action on $\overline{\mathcal{M}}_{\chi,n}^{\bullet}(\mathbb{P}^{1},\mu)$
        associated to $\Gamma$. Given a flag $(v,e)\in F(\Gamma)$, denote by $q_{(v,e)}\in C$ the node at
        which $C_{v}$ and $C_{e}$ intersect. Also let $\psi_{(v,e)}$ denote the first Chern class of the
        cotangent line bundles over $\overline{\mathcal{M}}_{\Gamma}$, i.e. the fiber at the fixed point (\ref{fixed_point})
        is given by $T^{*}_{q(v,e)}C_{v}$.
        The Euler class $e_{T}(\mathcal{N}_{\Gamma}^{vir})$ is given by;
        $$\frac{1}{e_{T}(\mathcal{N}_{\Gamma}^{vir})}=\frac{e_{T}(\hat{T}^{2})}{e_{T}(\hat{T}^{1})}$$
        where $\hat{V}$ denotes the moving part of any vector bundle $V$, and $T^{1}$, $T^{2}$ are the tangent
        space and the obstruction space of $\overline{\mathcal{M}}^{\bullet}_{\chi,n}(\mathbb{P}^{1},\mu)$, respectively.
        Here, $T^{1}$ and $T^{2}$ can be computed through the following two exact sequences \cite{Li2}:

            \begin{equation*}
                0\rightarrow\text{Ext}^{0}(\Omega_{C}(D),\mathcal{O}_{C})\rightarrow
                    H^{0}(\bf{D}\it^{\bullet})\rightarrow T^{1}\rightarrow
                    \text{Ext}^{1}(\Omega_{C}(D),\mathcal{O}_{C})\rightarrow
                    H^{1}(\bf{D}\it^{\bullet})\rightarrow T^{2}\rightarrow\rm 0\\
            \end{equation*}
            \begin{equation*}
                \xymatrix@R.1in{0\ar[r] & H^{0}(C,f^{*}(\omega_{\mathbb{P}^{1}[m]}(
                    \text{log}\,p_{1}^{(m)}))^{\vee})\ar[r] & H^{0}(\bf{D}\it^{\bullet})\ar[r]
                    & \bigoplus_{l=0}^{m-1}H^{0}_{et}(\bf{R}\it_{l}^{\bullet}) \\
                \ar[r]& H^{1}(C,f^{*}(\omega_{\mathbb{P}^{1}[m]}(\text{log}\,p_{1}^{(m)}))^{\vee})\ar[r] &H^{1}(
                    \bf{D}\it^{\bullet})\ar[r] & \bigoplus_{l=0}^{m-1}H^{1}_{et}(\bf{R}\it^{\bullet})\ar[r]& 0\\}\\
            \end{equation*}
            where $\omega_{\mathbb{P}^{1}[m]}$ is the dualizing sheaf of $\mathbb{P}^{1}[m]$,
            $D=x_{1}+\cdots+x_{l(\mu)}$ is the branch divisor, and
            for $n_{l}=\text{the number of nodes over }p_{1}^{(l)}$;
            \begin{align*}
                &H^{0}_{et}(\bf{R}\it_{l}^{\bullet})\cong\bigoplus_{q\in f^{-1}(p_{1}^{(l)})}
                    T_{q}\big(f^{-1}(\mathbb{P}^{1}_{(l)})\big)\,\,\cong\,\,\mathbb{C}^{\oplus n_{l}} \\
                &H^{1}_{et}(\bf{R}\it_{l}^{\bullet})\cong\big(T_{p_{1}^{(l)}}\mathbb{P}^{1}_{(l)}\otimes
                    T_{p_{1}^{(l)}}\mathbb{P}^{1}_{(l+1)}\big)^{\oplus(n_{l}-1)}
            \end{align*}
            Recall the map $\pi[m]:\mathbb{P}^{1}[m]\longrightarrow\mathbb{P}^{1}$, and observe that
            for $\hat{f}=\pi[m]\circ f$ we have
            \begin{equation*}\label{f_contracted}
                f^{*}\big(\omega_{\mathbb{P}^{1}[m]}(\text{log}\,p_{1}^{(m)})\big)^{\vee}
                \cong\hat{f}^{*}\mathcal{O}_{\mathbb{P}^{1}}(1)
            \end{equation*}
            Let $F_{\Gamma}$ be the set of fixed points associated to $\Gamma\in G_{\chi,n}(\mathbb{P}^{1},\mu)$
            and assume that
            \begin{equation*}
                \big[\,\,f:(C,x_{1},\cdots,x_{l(\mu)},z_{1},\cdots,z_{n})\longrightarrow\mathbb{P}^{1}[m]\,\,\big]
                \in\,F_{\Gamma}\subset\overline{\mathcal{M}}^{\bullet}_{\chi,n}(\mathbb{P}^{1},\mu)
            \end{equation*}
            The $\mathbb{C}^{*}$-action on $\overline{\mathcal{M}}_{\chi,n}^{\bullet}(\mathbb{P}^{1},\mu)$ induces
            $\mathbb{C}^{*}$-actions on
            \begin{align*}
                &\text{Ext}^{0}(\Omega_{C}(D),\mathcal{O}_{C})\quad,
                &&H^{0}(C,\hat{f}^{*}\mathcal{O}_{\mathbb{P}^{1}}(1))\quad,
                &&&\bigoplus_{l=0}^{m-1}H^{0}_{et}(\bf{R}\it_{l}^{\bullet}) \\
                &\text{Ext}^{1}(\Omega_{C}(D),\mathcal{O}_{C})\quad,
                &&H^{1}(C,\hat{f}^{*}\mathcal{O}_{\mathbb{P}^{1}}(1))\quad,
                &&&\bigoplus_{l=0}^{m-1}H^{1}_{et}(\bf{R}\it_{l}^{\bullet})
            \end{align*}
            The moving part of each of these groups form vector bundles over $\overline{\mathcal{M}}_{\Gamma}$.
            We will use the same notation $\hat{\cdot}$ to denote the induced vector bundles.
            In particular,
            \begin{align*}
                &\widehat{\bigoplus_{l=0}^{m-1}H^{0}_{et}(\bf{R}\it_{l}^{\bullet})}=0,\qquad\text{and}\\
                &\widehat{\bigoplus_{l=0}^{m-1}H^{1}_{et}(\bf{R}\it_{l}^{\bullet})}=
                \begin{cases}
                    0,&\text{if }m=0 \\
                    H^{1}_{et}(\bf{R}\it_{0}^{\bullet})=\big(T_{p_{1}^{(0)}}\mathbb{P}^{1}_{(0)}
                    \otimes T_{p_{1}^{(0)}}\mathbb{P}^{1}_{(1)}\big)^{\oplus(n_{0}-1)},&\text{if }m>0
                \end{cases}
            \end{align*}
            Hence we have;
            \begin{equation*}
                \frac{1}{e_{T}(\mathcal{N}_{\Gamma}^{vir})}=\frac{e_{T}(\hat{T}^{2})}{e_{T}(\hat{T}^{1})}
                =\frac{e_{T}\big(\widehat{\text{Ext}^{0}(\Omega_{C}(D),\mathcal{O}_{C})}\big)
                e_{T}\big(\widehat{H^{1}(C,\hat{f}^{*}\mathcal{O}_{\mathbb{P}^{1}}(1))}\big)
                e_{T}\big(\widehat{\bigoplus_{l=0}^{m-1}H^{1}_{et}(\bf{R}\it_{l}^{\bullet})}\big)}
                {e_{T}\big(\widehat{H^{0}(C,\hat{f}^{*}\mathcal{O}_{\mathbb{P}^{1}}(1))}\big)
                e_{T}\big(\widehat{\text{Ext}^{1}(\Omega_{C}(D),\mathcal{O}_{C})}\big)}
            \end{equation*}\\

            \underline{Case $m=0$:}\\
                For each $v\in V(\Gamma_{0})$ and $\mu_{v,1},\cdots,\mu_{v,l(\mu(v))}$ the ramification
                type in the vertex $v$, we have under the convention to write
                $\mu_{v,2}=\infty$ when $g(v)=0, l(\mu(v))=l(e(v))=1$;
                \begin{align*}
                    &\widehat{\bigoplus_{l=0}^{m-1}H^{1}_{et}(\bf{R}\it_{l}^{\bullet})_{v}}=0 \\
                    &\widehat{\text{Ext}^{0}(\Omega_{C}(D),\mathcal{O}_{C})}_{v}=
                    \begin{cases}
                        \Big(\frac{1}{\mu_{v,1}}\Big),&\text{if }v\in\text{I} \\
                        0,&\text{if }v\in\text{II or S}
                    \end{cases} \\
                    &\widehat{\text{Ext}^{1}(\Omega_{C}(D),\mathcal{O}_{C})}_{v}=
                    \begin{cases}
                        0,&\text{if }v\in\text{I}\\
                        \Big(\frac{1}{\mu_{v,1}}+\frac{1}{\mu_{v,2}}\Big),&\text{if }v\in\text{II}\\
                        \bigoplus_{i=1}^{l(\mu(v))}T_{q_{(v,e_{v,i})}}C_{v}\otimes
                            T_{q_{(v,e_{v,i})}}C_{e_{v,i}},&\text{if }v\in\text{S}
                    \end{cases}
                \end{align*}
                Hence we can compute their contributions to be;
                \begin{align*}
                    &e_{T}\big(\widehat{\bigoplus_{l=0}^{m-1}H^{1}_{et}(\bf{R}\it_{l}^{\bullet})_{v}}\big)=1\\
                    &e_{T}\big(\widehat{\text{Ext}^{0}(\Omega_{C}(D),\mathcal{O}_{C})}_{v}\big)=
                    \begin{cases}
                        \frac{u}{\mu_{v,1}},&\text{if }v\in\text{I}\\
                        1,&\text{if }v\in\text{II or S}
                    \end{cases} \\
                    &e_{T}\big(\widehat{\text{Ext}^{1}(\Omega_{C}(D),\mathcal{O}_{C})}_{v}\big)=
                    \begin{cases}
                        1,&\text{if }v\in\text{I}\\
                        \frac{u}{\mu_{v,1}}+\frac{u}{\mu_{v,2}},&\text{if }v\in\text{II}\\
                        \prod_{i=1}^{l(\mu(v))}\Big(\frac{u}{\mu_{v,i}}-\psi_{v,i}\Big),&\text{if }v\in\text{S}
                    \end{cases}
                \end{align*}
            For the contributions from the rest, consider the normalization sequence when $v\in$S;
            \begin{equation*}
                0\rightarrow\hat{f}^{*}\mathcal{O}_{\mathbb{P}^{1}}(1)\rightarrow
                \big(\hat{f}\mid_{C_{v}}\big)^{*}\mathcal{O}_{\mathbb{P}^{1}}(1)\oplus
                \bigoplus_{i=1}^{l(\mu(v))}\big(\hat{f}\mid_{C_{e_{v,i}}}\big)^{*}\mathcal{O}_{\mathbb{P}^{1}}(1)
                \rightarrow\bigoplus_{i=1}^{l(\mu(v))}\mathcal{O}_{\mathbb{P}^{1}}(1)_{p_{0}}\rightarrow 0
            \end{equation*}
            The corresponding long exact sequence becomes
            \begin{align*}\label{les_zero}
                0\longrightarrow &H^{0}\big(C,\hat{f}^{*}\mathcal{O}_{\mathbb{P}^{1}}(1)\big)\longrightarrow
                H^{0}\big(C_{v},(\hat{f}\mid_{C_{v}})^{*}\mathcal{O}_{\mathbb{P}^{1}}(1)\big)
                \oplus\bigoplus_{i=1}^{l(\mu(v))}H^{0}\big(C_{e_{v,i}},(\hat{f}\mid_{C_{e_{v,i}}})^{*}
                \mathcal{O}_{\mathbb{P}^{1}}(1)\big)\\
                &\longrightarrow\bigoplus_{i=1}^{l(\mu(v))}\mathcal{O}_{\mathbb{P}^{1}}(1)_{p_{0}}
                \longrightarrow H^{1}\big( C,\hat{f}^{*}\mathcal{O}_{\mathbb{P}^{1}}(1)\big) \\
                &\longrightarrow
                H^{1}\big(C_{v},(\hat{f}\mid_{C_{v}})^{*}\mathcal{O}_{\mathbb{P}^{1}}(1)\big)
                \oplus\bigoplus_{i=1}^{l(\mu(v))}H^{1}\big(C_{e_{v,i}},(\hat{f}\mid_{C_{e_{v,i}}})^{*}
                \mathcal{O}_{\mathbb{P}^{1}}(1)\big)\longrightarrow 0
            \end{align*}
            and the representations of $\mathbb{C}^{*}$ are given by
            \begin{align*}
                0\rightarrow &H^{0}\big(C,\hat{f}^{*}\mathcal{O}_{\mathbb{P}^{1}}(1)\big)\rightarrow
                H^{0}\big(C_{v},\mathcal{O}_{C_{v}}\big)\otimes(1)\oplus\bigoplus_{i=1}^{l(\mu(v))}
                \Big(\bigoplus_{a=1}^{\mu_{v,i}}\big(\frac{a}{\mu_{v,i}}\big)\Big)\rightarrow\\
                &\bigoplus_{i=1}^{l(\mu(v))}(1)\rightarrow H^{1}\big(C,\hat{f}^{*}
                \mathcal{O}_{\mathbb{P}^{1}}(1)\big)\rightarrow
                H^{1}\big(C_{v},\mathcal{O}_{C_{v}}\big)\otimes(1)\rightarrow 0
            \end{align*}
            Hence their ratio can computed as;
            \begin{equation*}
                \frac{e_{T}\big(\widehat{H^{1}(C,\hat{f}^{*}\mathcal{O}_{\mathbb{P}^{1}}(1))}\big)}
                {e_{T}\big(\widehat{H^{0}(C,\hat{f}^{*}\mathcal{O}_{\mathbb{P}^{1}}(1))}\big)}=
                \prod_{v}\Big[\Lambda_{g(v)}^{\vee}(u)u^{l(\mu(v))-1}\prod_{i=1}^{l(\mu(v))}
                \big(\frac{\mu_{v,i}^{\mu_{v,i}}}{\mu_{v,i}!}u^{-\mu_{v,i}}\big)\Big]
            \end{equation*}
            which also works for the case of $v\in$I or $v\in$II.\\

            \underline{Case $m>0$:}
                Let $\psi^{t}$ be the first Chern class of the line bundle over $\overline{\mathcal{M}}_{\Gamma}^{(1)}$ whose
                fiber at $\big[\,\,\hat{f}:\hat{C}\longrightarrow\mathbb{P}^{1}(m)\,\,\big]$ is $T_{p_{1}^{(0)}}^{*}\mathbb{P}^{1}(m)$.
                So $\psi^{t}=\nu_{v,i}\psi_{(v,e_{v,i})}$ for $v\in V(\Gamma)^{(1)},\,\,(v,e_{v,i})\in F$.
                By similar observation as in the case of $m=0$, we can find that
                \begin{align*}
                    &\widehat{\bigoplus_{l=0}^{m-1}H^{1}_{et}(\bf{R}\it_{l}^{\bullet})}=
                        \Big(T_{p_{1}^{(0)}}\mathbb{P}^{1}_{(0)}\otimes T_{p_{1}^{(0)}}\mathbb{P}^{1}_{(1)}\Big)^{\mid E(\Gamma)\mid-1} \\
                    &\widehat{\text{Ext}^{0}(\Omega_{C}(D),\mathcal{O}_{C})}_{v}=
                    \begin{cases}
                        \Big(\frac{1}{\nu_{v,1}}\Big),&\text{if }v\in\text{I} \\
                        0,&\text{if }v\in\text{II or S}
                    \end{cases} \\
                    &\widehat{\text{Ext}^{1}(\Omega_{C}(D),\mathcal{O}_{C})}_{v}=
                    \begin{cases}
                        0,&\text{if }v\in\text{I}\\
                        \Big(\frac{1}{\nu_{v,1}}+\frac{1}{\nu_{v,2}}\Big),&\text{if }v\in\text{II}\\
                        \bigoplus_{i=1}^{l(\nu(v))}T_{q_{(v,e_{v,i})}}C_{v}\otimes
                            T_{q_{(v,e_{v,i})}}C_{e_{v,i}},&\text{if }v\in\text{S or T}\\
                    \end{cases}
                \end{align*}
                Hence we can compute their contributions to be;
                \begin{align*}
                    &e_{T}\big(\widehat{\bigoplus_{l=0}^{m-1}H^{1}_{et}(\bf{R}\it_{l}^{\bullet})}\big)=
                        \big(-u-\psi^{t}\big)^{\mid E(\Gamma)\mid-1}\\
                    &e_{T}\big(\widehat{\text{Ext}^{0}(\Omega_{C}(D),\mathcal{O}_{C})}_{v}\big)=
                    \begin{cases}
                        \frac{u}{\nu_{v,1}},&\text{if }v\in\text{I}\\
                        1,&\text{if }v\in\text{II or S}
                    \end{cases} \\
                    &e_{T}\big(\widehat{\text{Ext}^{1}(\Omega_{C}(D),\mathcal{O}_{C})}_{v}\big)=
                    \begin{cases}
                        1,&\text{if }v\in\text{I}\\
                        \frac{u}{\nu_{v,1}}+\frac{u}{\nu_{v,2}},&\text{if }v\in\text{II}\\
                        \prod_{i=1}^{l(\nu(v))}\Big(\frac{u}{\nu_{v,i}}-\psi_{v,i}\Big),&\text{if }v\in\text{S}\\
                        \prod_{i=1}^{l(\nu(v))}\Big(\frac{-u}{\nu_{v,i}}-\psi_{v,i}\Big),&\text{if }v\in\text{T}\\
                    \end{cases}
                \end{align*}
            Also consider the following normalization sequence
            \begin{align*}
                0\rightarrow&\hat{f}^{*}\mathcal{O}_{\mathbb{P}^{1}}(1)\rightarrow
                \bigoplus_{S\cup T}\big(\hat{f}\mid_{C_{v}}\big)^{*}\mathcal{O}_{\mathbb{P}^{1}}(1)\oplus
                \bigoplus_{e\in E(\Gamma)}\big(\hat{f}\mid_{C_{e}}\big)^{*}\mathcal{O}_{\mathbb{P}^{1}}(1)\rightarrow\\
                &\bigoplus_{II}\mathcal{O}_{\mathbb{P}^{1}}(1)_{p_{0}}\oplus\bigoplus_{S}\Big(\bigoplus
                _{(v,e)\in F}\mathcal{O}_{\mathbb{P}^{1}}(1)_{p_{0}}\Big)\oplus\bigoplus_{T}\Big(\bigoplus
                _{(v,e)\in F}\mathcal{O}_{\mathbb{P}^{1}}(1)_{p_{1}}\Big)\rightarrow 0
            \end{align*}
            and the corresponding long exact sequence
            \begin{align*}
                0\rightarrow&H^{0}\big(C,\hat{f}^{*}\mathcal{O}_{\mathbb{P}^{1}}(1)\big)\rightarrow
                \bigoplus_{S\cup T}H^{0}\big(C_{v},\big(\hat{f}\mid_{C_{v}}\big)^{*}\mathcal{O}_{\mathbb{P}^{1}}(1)\big)
                \oplus\bigoplus_{e\in E(\Gamma)}H^{0}\big(C_{e},\big(\hat{f}\mid_{C_{e}}\big)^{*}\mathcal{O}_{\mathbb{P}^{1}}(1)\big)
                \rightarrow\\
                &\bigoplus_{II}\mathcal{O}_{\mathbb{P}^{1}}(1)_{p_{0}}\oplus\bigoplus_{S}\Big(\bigoplus
                _{(v,e)\in F}\mathcal{O}_{\mathbb{P}^{1}}(1)_{p_{0}}\Big)\oplus\bigoplus_{T}\Big(\bigoplus
                _{(v,e)\in F}\mathcal{O}_{\mathbb{P}^{1}}(1)_{p_{1}}\Big)\rightarrow H^{1}\big(C,\hat{f}^{*}\mathcal{O}_{\mathbb{P}^{1}}(1)\big)
                \rightarrow\\
                &\bigoplus_{ST}H^{1}\big(C_{v},(\hat{f}\mid_{C_{v}})^{*}\mathcal{O}_{\mathbb{P}^{1}}(1)\big)
                \oplus\bigoplus_{e\in E(\Gamma)}H^{1}\big( C_{e},(\hat{f}\mid_{C_{e}})^{*}\mathcal{O}_{\mathbb{P}^{1}}(1)\big)\rightarrow 0
            \end{align*}
            The representations of $\mathbb{C}^{*}$ are given by
            \begin{align*}
                0\rightarrow& H^{0}\big(C,\hat{f}^{*}\mathcal{O}_{\mathbb{P}^{1}}(1)\big)\rightarrow
                \bigoplus_{S}H^{0}\big(C_{v},\mathcal{O}_{C_{v}}\big)\otimes(1)\oplus\bigoplus_{T}
                H^{0}\big(C_{v},\mathcal{O}_{C_{v}}\big)\otimes(0)\oplus\bigoplus_{e\in E(\Gamma)}\Big(
                \bigoplus_{a=1}^{d(e)}\big(\frac{a}{d(e)}\big)\Big)\\
                &\rightarrow\bigoplus_{II}(1)\oplus\bigoplus_{S}\Big(\bigoplus_{(v,e)\in F}(1)\Big)\oplus
                \bigoplus_{T}\Big(\bigoplus_{(v,e)\in F}(0)\Big)\rightarrow H^{1}\big(C,\hat{f}^{*}
                \mathcal{O}_{\mathbb{P}^{1}}(1)\big)\rightarrow\\
                &\bigoplus_{S}H^{1}\big(C_{v},\mathcal{O}_{C_{v}}\big)\otimes(1)\oplus
                \bigoplus_{T}H^{1}\big(C_{v},\mathcal{O}_{C_{v}}\big)\otimes(0)\longrightarrow 0
            \end{align*}
            from which we can compute their ratio to be;
            \begin{equation*}
                \frac{e_{T}\big(\widehat{H^{1}(C,\hat{f}^{*}\mathcal{O}_{\mathbb{P}^{1}}(1))}\big)}
                {e_{T}\big(\widehat{H^{0}(C,\hat{f}^{*}\mathcal{O}_{\mathbb{P}^{1}}(1))}\big)}=
                \prod_{V(\Gamma)^{(0)}}\Big[\Lambda_{g(v)}^{\vee}(u)u^{l(\nu(v))-1}\Big]
                \prod_{i=1}^{l(\nu)}\Big[\frac{\nu_{i}^{\nu_{i}}}{\nu_{i}!}u^{-\nu_{i}}\Big]
            \end{equation*}\\
            
            After combining all the contributions, we find the following Feynman rules;
            \begin{align*}
                \frac{1}{e_{T}(\mathcal{N}_{\Gamma_{0}}^{vir})} =\qquad&
                    \Big[\prod_{i=1}^{l(\mu)}\frac{\mu_{i}^{\mu_{i}}}{\mu_{i}!}u^{-\mu_{i}}\Big]\times
                    \Big[\prod_{I}\frac{u}{\mu_{v,1}}\Big]\times
                    \Big[\prod_{II}\frac{u}{\frac{u}{\mu_{v,1}}+\frac{u}{\mu_{v,2}}}\Big]\\
                    &\times\Big[\prod_{S}\frac{\Lambda_{g(v)}^{\vee}(u)}{u}\Big(
                    \prod_{i=1}^{l(\mu(v))}\frac{u}{\frac{u}{\mu_{v,i}}-\psi_{v,i}}\Big)\Big]\\
                    \\
                \frac{1}{e_{T}(\mathcal{N}_{\Gamma}^{vir})} =\qquad&
                    \frac{-\prod\nu_{i}}{u+\psi^{t}}\times
                    \Big[\prod_{i=1}^{l(\nu)}\frac{\nu_{i}^{\nu_{i}}}{\nu_{i}!}u^{-\nu_{i}}\Big]\times
                    \Big[\prod_{I}\frac{u}{\nu_{v,1}}\Big]\times
                    \Big[\prod_{II}\frac{u}{\frac{u}{\nu_{n,1}}+\frac{u}{\nu_{v,2}}}\Big] \\
                    &\times\Big[\prod_{S}\frac{\Lambda_{g(v)}^{\vee}(u)}{u}\Big(
                    \prod_{i=1}^{l(\nu(v))}\frac{u}{\frac{u}{\nu_{v,i}}-\psi_{v,i}}\Big)\Big]\\
            \end{align*}
%-------------------------------------------------------------------------------
    \vskip 10mm
    \section{Double Hurwitz Numbers}
        \subsection{General Result of Hurwitz Numbers}
            Let $X$ be a Riemann surface of genus $h$. Given $n$ partitions $\eta^{1},\cdots,\eta^{n}$ of $d$
            , denote by $H^{X}_{d}(\eta^{1},\cdots,\eta^{n})^{\bullet}$ and $H^{X}_{d}(\eta^{1},\cdots,
            \eta^{n})^{\circ}$ the weighted counts of \it{possibly disconnected }\rm and \it{connected }
            \rm Hurwitz covers of type $(\eta^{1},\cdots,\eta^{n})$, respectively. The following
            \it{Burnside formula }\rm is well known: $$H^{X}_{d}(\eta^{1},\cdots,\eta^{n})^{\bullet}=\sum
            _{\mid\rho\mid=d}\Big(\frac{\text{dim}R_{\rho}}{d!}\Big)^{2-2h}\prod_{i=1}^{n}\mid C_{\eta^{i}}
            \mid\frac{\chi_{\rho}(C_{\eta^{i}})}{\text{dim}R_{\rho}}$$
        \subsection{Double Hurwitz Numbers}
            Consider a cover $C\longrightarrow\mathbb{P}^{1}$ of genus $g$, ramification type $\nu,\mu$ at
            two points $p_{0}$ and $p_{1}$ respectively, and ramification type $(2)$ at $r$ other points.
            By Riemann-Hurwitz formula, we have $r=2g-2+l(\nu)+l(\mu)$. Let $\eta^{1}=\cdots=\eta^{r}=(2)$
            and introduce notations $$H_{g}^{\circ}(\nu,\mu)=H^{\mathbb{P}^{1}}_{d}(\nu,\mu,\eta^{1},\cdots
            ,\eta^{r})^{\circ},\qquad H_{\chi}^{\bullet}(\nu,\mu)=H_{d}^{\mathbb{P}^{1}}(\nu,\mu,\eta^{1},
            \cdots,\eta^{r})^{\bullet}$$ By applying Burnside formula, we obtain
            $$H^{\bullet}_{\chi}(\nu,\mu) = \sum_{\mid\xi\mid=d}\Big(\mid C_{(2)}\mid\frac{\chi_{\xi}(C_{(2)}
            )}{\text{dim}R_{\xi}}\Big)^{r}\frac{\chi_{\xi}(C_{\nu})}{z_{\nu}}\frac{\chi_{\xi}(C_{\mu})}{z_{
            \mu}}   =\sum_{\mid\xi\mid=d} f_{\xi}(2)^{r}\frac{\chi_{\xi}(C_{\nu})}{z_{\nu}}\frac{\chi_{\xi}
            (C_{\mu})}{z_{\mu}}$$
            Define generating series of Double Hurwitz numbers as follows:
            \begin{align*}
                \Phi^{\circ}_{\nu,\mu}(\lambda) &=\sum_{g\geq 0}H^{\circ}_{g}(\nu,\mu)
                    \frac{\lambda^{2g-2+l(\nu)+\l(\mu)}}{(2g-2+l(\nu)+l(\mu))!}
                &\Phi^{\circ}(\lambda;p^{0},p^{\infty})=\sum_{\nu,\mu}\Phi^{\circ}_{\nu,\mu}(\lambda)
                    p^{0}_{\nu}p^{\infty}_{\mu} \\
                \Phi^{\bullet}_{\nu,\mu}(\lambda) &=\sum_{\chi}H^{\bullet}_{\chi}(\nu,\mu)
                    \frac{\lambda^{-\chi+l(\nu)+\l(\mu)}}{(-\chi+l(\nu)+l(\mu))!}
                &\Phi^{\bullet}(\lambda;p^{0},p^{\infty})=1+\sum_{\nu,\mu}\Phi^{\bullet}_{\nu,\mu}(\lambda)
                    p^{0}_{\nu}p^{\infty}_{\mu}
            \end{align*}
            We will need the following initial value formula of Double Hurwitz numbers;
            \begin{lemma}\label{lemma_double_initial}
                $$\Phi^{\bullet}_{\nu,\mu}(0)=\frac{1}{z_{\nu}}\delta_{\nu,\mu}$$
                \begin{proof}
                    This is a direct consequence of the orthogonal relations for characters of $S_{d}$
                    $$\sum_{\xi}\frac{\chi_{\nu}(C_{\xi})\chi_{\mu}(C_{\xi})}{z_{\xi}}=\delta_{\nu,\mu}
                    ,\qquad\text{and}\qquad \sum_{\mid\xi\mid=d}\chi_{\xi}(C_{\nu})\chi_{\xi}(C_{\mu})=
                    z_{\nu}\delta_{\nu,\mu}$$
                \end{proof}
            \end{lemma}

        \subsection{Relating Double Hurwitz Numbers with Hodge Integrals}
            We can extend the notion of $\mathbb{P}^{1}[m]$ to have bubble components on both directions.
            Denote by $$\mathbb{P}^{1}[m_{0},m_{\infty}]=\mathbb{P}^{1}_{(-m_{0})}\cup\cdots\cup\mathbb{P}^
            {1}_{(-1)}\cup\mathbb{P}^{1}_{(0)}\cup\mathbb{P}^{1}_{(1)}\cup\cdots\cup\mathbb{P}^{1}_{(m_{
            \infty})}$$
            As before, we call $\mathbb{P}^{1}_{(0)}$ the \it{root component }\rm, $\mathbb{P}^{1}_{0}(m_{
            0})=\mathbb{P}^{1}_{(-m_{0})}\cup\cdots\cup\mathbb{P}^{1}_{(-1)}$ the \it{bubble component at }
            \rm $0$, and $\mathbb{P}^{1}_{\infty}(m_{\infty})=\mathbb{P}^{1}_{(1)}\cup\cdots\cup\mathbb{P}^{1}_{
            (m_{\infty})}$ the \it{bubble component at }\rm $\infty$.
            Define $\overline{\mathcal{M}}_{\chi,n}^{\bullet}(\mu^{0},\mu^{\infty})$ as the moduli space of
            morphisms $f:C\longrightarrow\mathbb{P}^{1}[m_{0},m_{\infty}]$ such that
            \begin{itemize}
                \item[(1)] $\big(C;x_{1},\cdots,x_{l(\mu^{\infty})},y_{1},\cdots,y_{l(\mu^{0})}\big)$ is a
                    possibly-disconnected prestable curve with $l(\mu^{0})+l(\mu^{\infty})$ unordered
                    marked points.
                \item[(2)] $\chi=\sum_{i}(2-2g_{i})$ where $g_{i}$ is the genus of each connected component
                    of $C$.
                \item[(3)] As Cartier divisors,
                    $f^{-1}(p_{0}^{(-m_{0})})=\sum^{l(\mu^{0})}_{j=1}\mu_{j}^{0}y_{j},\quad
                    f^{-1}(p_{1}^{(m_{\infty})})=\sum^{l(\mu^{\infty})}_{i=1}\mu_{i}^{\infty}x_{i}$
                \item[(4)] The automorphism group of $f$ is finite. Here, an automorphism of
                    $f$ consists of an automorphism of the domain curve $C$ and automorphisms of the pointed
                    curves $(\mathbb{P}^{1}_{0}(m_{0}),p_{0}^{(-m_{0})},p_{1}^{(-1)})$ and $(\mathbb{P}^{1}
                    _{\infty}(m_{\infty}),p_{0}^{(1)},p_{1}^{(m_{\infty})})$, which is an element of
                    $(\mathbb{C}^{*})^{m_{0}}$ and $(\mathbb{C}^{*})^{m_{\infty}}$, respectively.
            \end{itemize}
            We can extend the standard action $t\cdot[z,w]=[tz,w]$ on $\mathbb{P}^{1}$ to $\mathbb{P}^{1}
            [m_{0},m_{\infty}]$ by trivial action on the bubble components at $0$ and $\infty$. Then this
            action induces $\mathbb{C}^{*}$-action on $\overline{\mathcal{M}}^{\bullet}_{\chi}(\mathbb{P}
            ^{1},\mu^{0},\mu^{\infty})$. Let $\pi[m_{0},m_{\infty}]:\mathbb{P}^{1}[m_{0},m_{\infty}]
            \longrightarrow\mathbb{P}^{1}$ be the projection which contracts both bubble components and
            $f_{r}=\pi[m_{0},m_{\infty}]\circ f$. Denote by $\nu$ the partition of $d=\vert\mu^{0}\vert=
            \vert\mu^{\infty}\vert$ by the degrees of $f_{r}$ on each rational irreducible components.
            For any morphism $f:\big(C,x_{i},y_{j}\big)\longrightarrow
            \mathbb{P}^{1}[m_{0},m_{\infty}]$ which represents a fixed point of $\overline{\mathcal{M}}
            ^{\bullet}_{\chi}(\mathbb{P}^{1},\mu^{0},\mu^{\infty})$ under this action, one of the following
            four cases must hold:
            \begin{itemize}
                \item{\underline{$m_{0}=m_{\infty}=0$}:} We have $f=f_{r}$, $\mu^{0}=\mu^{\infty}=\nu$
                \item{\underline{$m_{0}=0, m_{\infty}>0$}:} Let $\chi_{\infty}=\sum(2-2g_{i}^{\infty})$
                    where $g_{i}^{\infty}$ is the genus of each connected component of $f^{-1}(\mathbb{P}
                    ^{1}_{\infty}(m_{\infty}))$. In this case, we have $\mu^{0}=\nu,\chi_{\infty}=\chi,
                    \chi_{0}=2l(\nu)$
                \item{\underline{$m_{0}>0, m_{\infty}=0$}:} Let $\chi_{0}=\sum(2-2g_{j}^{0})$ where
                    $g_{j}^{0}$ is the genus of each connected component of $f^{-1}(\mathbb{P}^{1}_{0}
                    (m_{0}))$. In this case, we have $\mu^{\infty}=\nu, \chi_{0}=\chi,\chi_{\infty}=2l(\nu)$.
                \item{\underline{$m_{0}>0, m_{\infty}>0$}:} We have $\chi=\chi_{0}+\chi_{\infty}-2l(\nu)$.
            \end{itemize}
            Hence, we can see that $\chi_{0},\chi_{\infty},\nu$ determines each connected component of
            $\overline{\mathcal{M}}^{\bullet}_{\chi}(\mathbb{P}^{1},\mu^{0},\mu^{\infty})^{\mathbb{C}^{*}}$.
            Consider the branch morphism
            $$\text{Br}:\overline{\mathcal{M}}_{\chi}^{\bullet}(\mathbb{P}^{1},\mu^{0},\mu^{\infty})
            \longrightarrow\text{Sym}^{-\chi+l(\mu^{0})+l(\mu^{\infty})}\mathbb{P}^{1}\,\cong\,\mathbb{P}
            ^{-\chi+l(\mu^{0})+l(\mu^{\infty})}$$
            The Double Hurwitz numbers for possibly disconnected covers of $\mathbb{P}^{1}$ can be defined by
            $$H^{\bullet}_{\chi}(\mu^{0},\mu^{\infty})=\frac{1}{\mid\text{Aut}(\mu^{0})\mid\mid\text{Aut}
            (\mu^{\infty})\mid}\int_{\big[\overline{\mathcal{M}}_{\chi}^{\bullet}(\mu^{0},\mu^{\infty})\big]
            ^{vir}}\text{Br}^{*}\big(H^{-\chi+l(\mu^{0})+l(\mu^{\infty})}\big)$$
            under the assumption $-\chi+l(\mu^{0})+l(\mu^{\infty})>0$
            where $H\in H^{2}(\mathbb{P}^{-\chi+l(\mu^{0})+l(\mu^{\infty})})$ is the hyperplane class.
            We want to compute this integration by virtual localization.
            The connected components of $\overline{\mathcal{M}}^{\bullet}_{\chi}(\mathbb{P}
            ^{1},\mu^{0},\mu^{\infty})^{\mathbb{C}^{*}}$ can be described as follows:
            \begin{align*}
                &\bullet\underline{m_{0}=0,m_{\infty}>0}: \mathcal{F}(\nu;2l(\nu),\chi)\cong\Big(\overline{
                    \mathcal{M}}^{\bullet}_{\chi}(\mathbb{P}^{1},\nu,\mu^{\infty})//\mathbb{C}^{*}\Big)/
                    \prod\mathbb{Z}_{\mu_{i}^{\infty}}\\
                &\bullet\underline{m_{0}>0,m_{\infty}=0}: \mathcal{F}(\nu;\chi,2l(\nu))\cong\Big(\overline{
                    \mathcal{M}}^{\bullet}_{\chi}(\mathbb{P}^{1},\mu^{0},\nu)//\mathbb{C}^{*}\Big)/
                    \prod\mathbb{Z}_{\mu_{j}^{0}}\\
                &\bullet\underline{m_{0}>0,m_{\infty}>0}:\\
                &\qquad\mathcal{F}(\nu;\chi_{0},\chi_{\infty})\cong\Big(
                    \big(\overline{\mathcal{M}}_{\chi_{0}}^{\bullet}(\mathbb{P}^{1},\mu^{0},\nu)//\mathbb{C}
                    ^{*}\big)\times\big(\overline{\mathcal{M}}_{\chi_{\infty}}^{\bullet}(\mathbb{P}^{1},
                    \nu,\mu^{\infty})//\mathbb{C}^{*}\big)\Big)
                    /\big(\text{Aut}(\nu)\prod_{i=1}^{l(\nu)}
                    \mathbb{Z}_{\nu_{i}}\big)
            \end{align*}
            Let $\mathcal{N}^{vir}_{\nu;\chi_{0},\chi_{\infty}}$ be the pull-back of the virtual normal
            bundle of $\mathcal{F}(\nu,\chi_{0},\chi_{\infty})$ in $\overline{\mathcal{M}}^{\bullet}_{\chi}
            (\mathbb{P}^{1},\mu^{0},\mu^{\infty})$. By computations similar to those $e_{T}(\mathcal{N}_{\Gamma}^{vir})$,
            we obtain
            $$\frac{1}{e_{T}\big(\mathcal{N}^{vir}_{\nu;2l(\nu),\chi}\big)}=\frac{-a_{\nu}}{u+\psi^{0}},
            \qquad\frac{1}{e_{T}\big(\mathcal{N}^{vir}_{\nu;\chi,2l(\nu)}\big)}=\frac{a_{\nu}}{u-\psi^{
            \infty}},\qquad\frac{1}{e_{T}\big(\mathcal{N}^{vir}_{\nu;\chi_{0},\chi_{\infty}}\big)}=\frac{
            -A_{\nu}}{u+\psi^{0}}\times\frac{a_{\nu}}{u-\psi^{\infty}}$$
            where $\psi^{0}$ and $\psi^{\infty}$ are the first Chern classes of the cotangent line bundle
            $T^{*}_{p_{0}^{(0)}}\mathbb{P}^{1}[m_{0},m_{\infty}]$ and $T^{*}_{p_{1}^{(0)}}\mathbb{P}^{1}
            [m_{0},m_{\infty}]$, respectively. Let $r=-\chi+l(\mu^{0})+l(\mu^{\infty})$ and observe that
            \begin{align*}
                &\text{Br}(\mathcal{F}(\nu;\chi_{0},\chi_{\infty}))=\big(-\chi_{\infty}+l(\nu)+l(\mu^{
                    \infty})\big)p_{1}+\big(-\chi_{0}+l(\mu^{0})+l(\nu)\big)p_{0}\quad\in\,\,\mathbb{P}^{r} \\
                &i^{*}_{\nu,\chi_{0},\chi_{\infty}}\text{Br}^{*}\Big(\prod_{k=1}^{r}\big(H-w_{k}\big)\Big)=
                    \Big(\prod_{k=1}^{r}\big(-\chi_{0}+l(\mu^{0})+l(\nu)-w_{k}\big)\Big)u^{r}
            \end{align*}
            By taking special values for $w$:
            $$w=(0,1,\cdots,-\chi+l(\mu^{0})+l(\mu^{\infty})-1)\quad\text{and}\quad
            w=(1,2,\cdots,-\chi+l(\mu^{0})+l(\mu^{\infty}))$$
            , we can compute to obtain
            \begin{align*}
                \frac{H^{\bullet}_{\chi}(\mu^{0},\mu^{\infty})}{(-\chi+l(\mu^{0})+l(\mu^{\infty}))!}
                =&\frac{1}{\mid\text{Aut}(\mu^{0})\mid\mid\text{Aut}(\mu^{\infty})\mid}\int_{\big[
                    \overline{\mathcal{M}}^{\bullet}_{\chi}(\mathbb{P}^{1},\mu^{0},\mu^{\infty})//\mathbb{C}
                    ^{*}\big]^{vir}}\big(\psi^{0}\big)^{-\chi+l(\mu^{0})+l(\mu^{\infty})-1} \\
                =&\frac{1}{\mid\text{Aut}(\mu^{0})\mid\mid\text{Aut}(\mu^{\infty})\mid}\int_{\big[
                    \overline{\mathcal{M}}^{\bullet}_{\chi}(\mathbb{P}^{1},\mu^{0},\mu^{\infty})//\mathbb{C}
                    ^{*}\big]^{vir}}\big(\psi^{\infty}\big)^{-\chi+l(\mu^{0})+l(\mu^{\infty})-1} \\
            \end{align*}
%-------------------------------------------------------------------------------
    \vskip 10mm
    \section{Recursion Formula}
            In this section, I will summarize the results from previous sections to prove the following
            recursion formula on Hodge integrals. Let $e=(k_{1},\cdots,k_{n})$ be a partition where
            $k_{i}$'s are allowed to be zero.
            \begin{theorem}
                For any partition $\mu$ and $e$ with $\vert e\vert<\vert\mu\vert+l(\mu)-\chi$, we have
                \begin{equation}\label{formula_recursion}
                    \big[\lambda^{l(\mu)-\chi}\big]\sum_{\mid\nu\mid=\mid\mu\mid}\Phi^{\bullet}_{\mu,\nu}(-\lambda)
                    z_{\nu}\mathcal{D}_{\nu,e}^{\bullet}(\lambda)=0
                \end{equation}
                where the sum is taken over all partitions $\nu$ of the same size as $\mu$.
            \end{theorem}
            Here $\big[\lambda^{a}\big]$ means taking the coefficient of $\lambda^{a}$.
            Let me first introduce some notations;
            \begin{align*}
                \mathcal{D}_{g,\nu,e}=\quad&
                    \frac{\nu_{1}^{\nu_{1}-2}}{\nu_{1}!}
                    \qquad\qquad\qquad\qquad\qquad\qquad\qquad\qquad
                    ,\text{if }\big(g,l(\nu)+l(e)\big)=\big(0,1\big)\\
                    &\frac{1}{\vert\text{Aut }\nu\vert}\frac{\nu_{1}^{\nu_{1}}\nu_{2}^{\nu_{2}}}
                    {\nu_{1}!\nu_{2}!}\frac{1}{\nu_{1}+\nu_{2}}
                    \qquad\qquad\qquad\qquad\quad,\text{if }\big(g,l(\nu),l(e)\big)=\big(0,2,0\big)\\
                    &\frac{\nu_{1}^{\nu_{1}}}{\nu_{1}!}
                    \sum_{k=0}^{e_{1}}\frac{1}{\nu_{1}^{1+k}}{e_{1}\choose k}
                    \qquad\qquad\qquad\qquad\qquad\quad,\text{if }\big(g,l(\nu),l(e)\big)=\big(0,1,1\big)\\
                    &\frac{1}{l(e)!\,\mid\text{Aut }\nu\mid}
                        \Big[\prod_{i=1}^{l(\nu)}\frac{\nu_{i}^{\nu_{i}}}{\nu_{i}!}\Big]
                        \int_{\overline{\mathcal{M}}_{g,l(\nu)+l(e)}}
                        \frac{\Lambda_{g}^{\vee}(1)\prod_{j=1}^{l(e)}(1-\psi_{j})^{e_{j}}}
                        {\prod_{i=1}^{l(\nu)}\big(1-\nu_{i}\psi_{i}\big)}
                        \quad,\text{otherwise}\\
                \mathcal{D}(\lambda,p,q)=\quad&\sum_{\mid\nu\mid\geq 1}\sum_{g\geq 0}\lambda^{2g-2+l(\nu)}p_{\nu}q_{e}\mathcal{D}_{g,\nu}\\
                \mathcal{D}^{\bullet}(\lambda,p,q)=\quad&\text{exp}\big(\mathcal{D}(\lambda,p,q)\big)
                =:\sum_{\mid\nu\mid\geq 0}\lambda^{-\chi+l(\nu)}p_{\nu}q_{e}\mathcal{D}^{\bullet}_{\chi,\nu,e}
                =\sum_{\mid\nu\mid\geq 0}p_{\nu}q_{e}\mathcal{D}^{\bullet}_{\nu}(\lambda)
            \end{align*}
            where $p_{i}$, $q_{j}$'s are formal variables with $p_{\nu}=p_{\nu_{1}}\times\cdots\times p_{\nu_{l(\nu)}}$
            and $q_{e}=q_{e_{1}}\times\cdots\times q_{e_{l(e)}}$.
            \begin{proof}
                For any given $\mu$ and $\chi$ such that $\vert\mu\vert+l(\mu)>\chi$,
                applying the localization formula to the class $\prod_{j=1}^{n}\psi_{j}^{k_{j}}\text{ev}_{j}^{*}H$
                where $H$ is the hyperplane class of $\mathbb{P}^{1}$ yields;
                \begin{align*}
                0=&\int_{\overline{\mathcal{M}}_{\chi,n}(\mathbb{P}^{1},\mu)}\prod_{j=1}^{n}\psi_{j}^{k_{j}}\text{ev}_{j}^{*}H
                    \qquad\qquad\text{since}\quad\text{deg}\prod_{j=1}^{n}\psi_{j}^{k_{j}}\text{ev}_{j}^{*}H<\text{dim}\overline{\mathcal{M}}_{\chi,n}(\mathbb{P}^{1},\mu)\\
                    =&\sum_{\Gamma_{0}\in G_{\chi,n}^{0}(\mathbb{P}^{1},\mu)}
                    \frac{1}{\vert A_{\Gamma_{0}}\vert}\int_{\overline{
                    \mathcal{M}}_{\Gamma_{0}}}\frac{\prod_{j=1}^{n}(u-\psi_{j})^{k_{j}}\text{ev}_{j}^{*}H_{T}}
                    {e_{T}(\mathcal{N}_{\Gamma_{0}}^{vir})}\\
                    &+
                    \sum_{\Gamma\in G_{\chi,n}^{\infty}(\mathbb{P}^{1},\mu)}\frac{1}{\vert A_{\Gamma}\vert}
                    \int_{\overline{\mathcal{M}}_{\Gamma}}\frac{\prod_{j=1}^{n}(u-\psi_{j})^{k_{j}}\text{ev}_{j}^{*}H_{T}}
                    {e_{T}(\mathcal{N}_{\Gamma}^{vir})}
                \end{align*}
                Here $H_{T}$ is the lift of $H$ to the equivariant hyperplane class.
                Choose $H$ in such a way that $H(0)=0$ and $H(\infty)=u$, then we have $H_{T}(0)=u$ and $H_{T}(\infty)=0$.
                Also let $u=1$ for simplicity, then the formula reduces to:
                \begin{equation*}
                    \sum_{\Gamma_{0}\in\overline{G}_{\chi,n}^{0}(\mathbb{P}^{1},\mu)}
                    \frac{1}{\vert A_{\Gamma_{0}}\vert}\int_{\overline{
                    \mathcal{M}}_{\Gamma_{0}}}\frac{\prod_{j=1}^{n}(1-\psi_{j})^{k_{j}}}
                    {e_{T}(\mathcal{N}_{\Gamma_{0}}^{vir})} +
                    \sum_{\Gamma\in\overline{G}_{\chi,n}^{\infty}(\mathbb{P}^{1},\mu)}\frac{1}{\vert A_{\Gamma}\vert}
                    \int_{\overline{\mathcal{M}}_{\Gamma}}\frac{\prod_{j=1}^{n}(1-\psi_{j})^{k_{j}}}
                    {e_{T}(\mathcal{N}_{\Gamma}^{vir})}
                    =0
                \end{equation*}
                where $\overline{G}_{\chi,n}^{0}$ and $\overline{G}_{\chi,n}$ are the set of graphs corresponding to
                fixed locus with $m=0$ and $m>0$ with all the marked points $z_{1},\cdots,z_{n}$ concentrated on the
                vertices in $V(\Gamma_{0})^{(0)}$ and $V(\Gamma)^{(0)}$, respectively.
                We can compute the summand for $\Gamma_{0}$ as follows:
                \begin{align*}
                    \int_{\overline{\mathcal{M}}_{\Gamma_{0}}}&\frac{\prod_{j=1}^{n}(1-\psi_{j})^{k_{j}}}
                    {e_{T}(\mathcal{N}_{\Gamma_{0}}^{vir})} =
                    \Big[\prod_{i=1}^{l(\mu)}\frac{\mu_{i}^{\mu_{i}}}{\mu_{i}!}\Big]\times
                    \Big[\prod_{I}\int_{\{pt\}}\frac{1}{\mu_{v,1}}\Big]\times
                    \Big[\prod_{II}\int_{\{pt\}}\frac{1}{\frac{1}{\mu_{v,1}}+\frac{1}{\mu_{v,2}}}\Big]\\
                    &\times\Big[\prod_{S}\int_{\overline{\mathcal{M}}_{g(v),l(\mu(v))+l(e(v))}}
                    \frac{\Lambda_{g(v)}^{\vee}(1)\prod_{j=1}^{j(v)}(1-\psi_{v,j})^{k_{v,j}}}
                    {\prod_{i=1}^{l(\mu(v))}\big(\frac{1}{\mu_{v,i}}-\psi_{v,i}\big)}\Big]\\
                    =&\Big[\prod_{I}\mu_{v,1}\frac{\mu_{v,1}^{\mu_{v,1}-2}}{\mu_{v,1}!}\Big]\times
                    \Big[\prod_{II}\mu_{v,1}\mu_{v,2}\frac{\mu_{v,1}^{\mu_{v,1}}\mu_{v,2}^{\mu_{v,2}}}
                    {\mu_{v,1}!\mu_{v,2}!}\frac{1}{\mu_{v,1}+\mu_{v,2}}\Big]\\
                    &\times\Big[\prod_{S}\big(\prod_{i=1}^{l(\mu(v))}\mu_{v,i}\big)
                    \Big(\prod_{i=1}^{l(\mu(v))}\frac{\mu_{v,i}^{\mu_{v,i}}}{\mu_{v,i}!}\Big)
                    \int_{\overline{\mathcal{M}}_{g(v),l(\mu(v))+l(e(v))}}
                    \frac{\Lambda_{g(v)}^{\vee}(1)\prod_{j=1}^{j(v)}(1-\psi_{v,j})^{k_{v,j}}}
                    {\prod_{i=1}^{l(\mu(v))}\big(1-\mu_{v,i}\psi_{v,i}\big)}\Big]\\
                    =&\prod_{V(\Gamma_{0})}z_{\mu(v)}j(v)!\mathcal{D}_{g(v),\mu(v)}
                \end{align*}
                Similarly, we can compute the summand for $\Gamma$ as follows:
                \begin{align*}
                    \int_{\overline{\mathcal{M}}_{\Gamma}}&\frac{\prod_{j=1}^{n}(1-\psi_{j})^{k_{j}}}
                    {e_{T}(\mathcal{N}_{\Gamma}^{vir})} =
                    \Big[\prod_{i=1}^{l(\nu)}\frac{\nu_{i}^{\nu_{i}}}{\nu_{i}!}\Big]\times
                    \Big[\prod_{I}\int_{\{pt\}}\frac{1}{\nu_{v,1}}\Big]\times
                    \Big[\prod_{II}\int_{\{pt\}}\frac{1}{\frac{1}{\nu_{n,1}}+\frac{1}{\nu_{v,2}}}\Big] \\
                    &\times\Big[\prod_{S}\int_{\overline{\mathcal{M}}_{g(v),l(\nu(v))+l(e(v))}}
                    \frac{\Lambda_{g(v)}^{\vee}(1)\prod_{j=1}^{j(v)}(1-\psi_{v,j})^{k_{v,j}}}
                    {\prod_{i=1}^{l(\nu(v))}\big(\frac{1}{\nu_{v,i}}-\psi_{v,i}\big)}\Big]
                    \times\Big[\int_{\overline{\mathcal{M}}^{(1)}_{\Gamma}}\frac{-\prod\nu_{i}}{1+\psi^{t}}\Big]\\
                    =&\Big[\prod_{I}\nu_{v,1}\frac{\nu_{v,1}^{\nu_{v,1}-2}}{\nu_{v,1}!}\Big]\times
                    \Big[\prod_{II}\nu_{v,1}\nu_{v,2}\frac{\nu_{v,1}^{\nu_{v,1}}\nu_{v,2}^{\nu_{v,2}}}
                    {\nu_{v,1}!\nu_{v,2}!}\frac{1}{\nu_{v,1}+\nu_{v,2}}\Big]\\
                    &\times\Big[\prod_{S}\big(\prod_{i=1}^{l(\nu(v))}\nu_{v,i}\big)
                    \Big(\prod_{i=1}^{l(\nu(v))}\frac{\nu_{v,i}^{\nu_{v,i}}}{\nu_{v,i}!}\Big)
                    \int_{\overline{\mathcal{M}}_{g(v),l(\nu(v))+l(e(v))}}
                    \frac{\Lambda_{g(v)}^{\vee}(1)\prod_{j=1}^{j(v)}(1-\psi_{v,j})^{k_{v,j}}}
                    {\prod_{i=1}^{l(\nu(v))}\big(1-\nu_{v,i}\psi_{v,i}\big)}\Big]\\
                    &\times\Big[(-1)^{-\chi+l(\mu)+l(\nu)}\prod_{i=1}^{l(\nu)}\nu_{i}\Big]
                    \int_{\overline{\mathcal{M}}_{\Gamma}^{(1)}}\big(\psi^{t}\big)^{-\chi_{\infty}+l(\mu)+l(\nu)-1} \\
                    =&\Big[\prod_{V(\Gamma_{0})}z_{\nu(v)}\,j(v)!\,\mathcal{D}_{g(v),\nu(v),e(v)}\Big]
                    \times\Big[(-1)^{-\chi+l(\mu)+l(\nu)}\prod_{i=1}^{l(\nu)}\nu_{i}\Big]
                    \int_{\big[\overline{\mathcal{M}}_{\Gamma}^{(1)}\big]^{vir}}\big(\psi^{t}\big)^{-\chi_{\infty}+l(\mu)+l(\nu)-1}
                \end{align*}
            And the integration over $\overline{\mathcal{M}}_{\Gamma}^{(1)}$ can be related to
            Double Hurwitz Numbers as follows:
            \begin{align*}
                \vert\text{Aut }\mu\vert\,\vert\text{Aut }\nu\vert\,\,&H^{\bullet}_{\chi_{\infty}}(\mu,\nu)=
                \big(-\chi_{\infty}+l(\mu)+l(\nu)\big)!\,\,
                \int_{[\overline{\mathcal{M}}^{\bullet}_{\chi_{\infty}}(\mathbb{P}^{1},\mu,\nu)//
                    \mathbb{C}^{*}]^{vir}}(\psi^{0})^{-\chi_{\infty}+l(\mu)+l(\nu)-1}\\
                =&\frac{\big(-\chi_{\infty}+l(\mu)+l(\nu)\big)!}
                {\Big(\prod n_{k}!\Big)\Big(\prod_{V(\Gamma)^{(1)}}\vert\text{Aut }\mu(v)\vert\,\,
                    \vert\text{Aut }\nu(v)\vert\Big)}\int_{\big[\overline{\mathcal{M}}_{\Gamma}^{(1)}\big]^{vir}}
                    \big(\psi^{t}\big)^{-\chi_{\infty}+l(\mu)+l(\nu)-1}
            \end{align*}
            since marked points in $\overline{\mathcal{M}}_{\Gamma}^{(1)}$ are ordered.
            Recall that $\vert A_{\Gamma_{0}}\vert$ and $\vert A_{\Gamma}\vert$ are given by:
            \begin{align*}
                \vert A_{\Gamma_{0}}\vert=&\Big(\prod_{i=1}^{l(\mu)}\mu_{i}\Big)\prod_{k}m_{k}!\,\,
				\big(j(v_{k})!\,\vert\,\text{Aut }\mu(v_{k})\,\vert\big)^{m_{k}} \\
                \vert A_{\Gamma}\vert=&\Big(\prod_{i=1}^{l(\nu)}\nu_{i}\Big)\Big(\prod_{k}m_{k}!\,\,
				\big(j(v_{k})!\,\vert\,\text{Aut }\nu(v_{k})\,\vert\big)^{m_{k}}\Big)\vert\text{Aut }
				\mu\,\vert\\
		        &\times\Big(\prod n_{k}!\Big)\Big(\prod_{V(\Gamma)^{(1)}}\,j(v)!\,\vert\,\text{Aut }
				\mu(v)\vert\,\,\vert\text{Aut }\nu(v)\vert\Big)
            \end{align*}
            Also observe that
            \begin{align*}
                \frac{\prod_{V(\Gamma_{0})}z_{\mu(v)}\,j(v)!\,\mathcal{D}_{g(v),\mu(v),e(v)}}
                {\Big(\prod_{i=1}^{l(\mu)}\mu_{i}\Big)\prod_{k}m_{k}!\,\big(\,j(v)!\,
                \vert\,\text{Aut }\mu(v_{k})\,\vert\big)^{m_{k}}}=&
                \frac{1}{\vert V(\Gamma_{0})\vert!}{\vert V(\Gamma_{0})\vert\choose{m_{1},\cdots,m_{l}}}
                \prod_{k}\mathcal{D}_{g(v_{k}),\mu(v_{k}),e(v_{k})}
            \end{align*}
            which is the coefficient of $\lambda^{-\chi+l(\mu)}p_{\mu}q_{e}$ in the expansion of
            $\mathcal{D}^{\bullet}(\lambda,p,q)$.
            Now the original equation can be simplified as follows;
            \begin{align*}
                0=&\sum_{\Gamma_{0}\in\overline{G}_{\chi,n}^{0}(\mathbb{P}^{1},\mu)}
                    \frac{1}{\vert A_{\Gamma_{0}}\vert}
                    \int_{\overline{\mathcal{M}}_{\Gamma_{0}}}
                    \frac{\prod_{j=1}^{n}(1-\psi_{j})^{k_{j}}}{e_{T}(\mathcal{N}_{\Gamma_{0}}^{vir})} +
                    \sum_{\Gamma\in\overline{G}_{\chi,n}^{\infty}(\mathbb{P}^{1},\mu)}
                    \frac{1}{\vert A_{\Gamma}\vert}
                    \int_{\overline{\mathcal{M}}_{\Gamma}}
                    \frac{\prod_{j=1}^{n}(1-\psi_{j})^{k_{j}}}{e_{T}(\mathcal{N}_{\Gamma}^{vir})} \\
                =&\sum_{\Gamma_{0}\in\overline{G}_{\chi,n}^{0}(\mathbb{P}^{1},\mu)}
                    \frac{1}{\vert V(\Gamma_{0})\vert!}{\vert V(\Gamma_{0})\vert\choose{m_{1},\cdots,m_{l}}}
                    \prod_{k}\mathcal{D}_{g(v_{k}),\mu(v_{k}),e(v_{k})} \\
                    &+\sum_{\Gamma\in\overline{G}_{\chi,n}^{\infty}(\mathbb{P}^{1},\mu)}\Big[
                    \frac{1}{\vert V(\Gamma)^{(0)}\vert!}{\vert V(\Gamma)^{(0)}\vert\choose{m_{1},\cdots,m_{l}}}
                    \prod_{k}\mathcal{D}_{g(v_{k}),\nu(v_{k}),e(v_{k})}\\
                    &\times\frac{(-1)^{-\chi_{\infty}+l(\mu)+l(\nu)}\prod_{i=1}^{l(\nu)}\nu_{i}}{\vert\text{Aut }\mu\vert\,
                    \Big(\prod n_{k}!\Big)\Big(\prod_{V(\Gamma)^{(1)}}\vert\text{Aut }\mu(v)\vert\,\,
                    \vert\text{Aut }\nu(v)\vert\Big)}\int_{\big[\overline{\mathcal{M}}_{\Gamma}^{(1)}\big]^{vir}}
                    \big(\psi^{t}\big)^{-\chi_{\infty}+l(\mu)+l(\nu)-1}\Big]\\
                =&\sum_{\Gamma_{0}\in\overline{G}_{\chi,n}^{0}(\mathbb{P}^{1},\mu)}
                    \frac{1}{\vert V(\Gamma_{0})\vert!}{\vert V(\Gamma_{0})\vert\choose{m_{1},\cdots,m_{l}}}
                    \prod_{k}\mathcal{D}_{g(v_{k}),\mu(v_{k}),e(v_{k})}\\ 
                    &+\sum_{\Gamma\in\overline{G}_{\chi,n}^{\infty}(\mathbb{P}^{1},\mu)}\Big[
                    \frac{1}{\vert V(\Gamma)^{(0)}\vert!}{\vert V(\Gamma)^{(0)}\vert\choose{m_{1},\cdots,m_{l}}}
                    \prod_{k}\mathcal{D}_{g(v_{k}),\nu(v_{k}),e(v_{k})}\\
                    &\times\frac{(-1)^{-\chi_{\infty}+l(\mu)+l(\nu)}z_{\nu}}{\vert\text{Aut }\mu\vert\,\,\vert
                    \text{Aut }\nu\vert}\int_{[\overline{\mathcal{M}}
                    ^{\bullet}_{\chi_{\infty}}(\mathbb{P}^{1},\mu,\nu)//\mathbb{C}^{*}]^{vir}}(\psi^{0})^{
                    -\chi_{\infty}+l(\mu)+l(\nu)-1}\Big]\\
                =&\mathcal{D}^{\bullet}_{\chi,\mu,e} + \sum_{\nu}\sum_{-\chi_{\infty}+l(\mu)+l(\nu)\neq 0}
                    \frac{(-1)^{-\chi_{\infty}+l(\mu)+l(\nu)}H_{\chi_{\infty}}^{\bullet}(\mu,\nu)}
                    {(-\chi_{\infty}+l(\mu)+l(\nu))!}z_{\nu}\mathcal{D}^{\bullet}_{\chi_{0},\nu,e} \\
                =&\sum_{\nu}\sum_{\chi_{0},\chi_{\infty}}
                    \frac{(-1)^{-\chi_{\infty}+l(\mu)+l(\nu)}H_{\chi_{\infty}}^{\bullet}(\mu,\nu)}
                    {(-\chi_{\infty}+l(\mu)+l(\nu))!}z_{\nu}\mathcal{D}^{\bullet}_{\chi_{0},\nu,e}
            \end{align*}
            where $\chi=\chi_{0}+\chi_{\infty}-2\,l(\nu)$ and the initial value formula for Double Hurwitz numbers
            is used at the last equality. Summing over $\chi$ yields that we have for all $\vert e\vert+\chi <\vert\mu\vert+l(\mu)$
            \begin{equation*}
                \big[\lambda^{l(\mu)-\chi}\big]
                \sum_{\vert\nu\vert=\vert\mu\vert}\Phi_{\mu,\nu}^{\bullet}(-\lambda)
                z_{\nu}\mathcal{D}^{\bullet}_{\nu,e}(\lambda)=0
            \end{equation*}
        \end{proof}
%-------------------------------------------------------------------------------
    \vskip 10mm
    \section{Computing Hodge Integrals with one $\lambda$-class}
        It is enough to consider the case of $\mu=(d)$ for some positive integers $d$ to compute all Hodge integrals with
        at-most one $\lambda$-class. And in this case, we have a closed formula for
        the Double Hurwitz Numbers as follows;
        \begin{theorem}\rm{( \cite{Gou-Jac-Vak}, Theorem 3.1.) }\it
            Let $r=r^{g}_{(d),\beta}$. For $g\geq 0$, and $\beta\vdash d$ with $n$ parts,
            \begin{align*}
                H^{g}_{(d),\beta}
                =&r!d^{r-1}\big[t^{2g}\big]\prod_{k\geq 1}\Big(\frac{\rm{sinh}\it(kt/2)}{kt/2}\Big)^{c_{k}}
                =\frac{r!d^{r-1}}{2^{2g}}\sum_{\lambda\vdash g}\frac{\xi_{2\lambda}S_{2\lambda}}{\vert\text{Aut }\lambda\vert}
            \end{align*}
        where $r=2g-1+l(\beta)$ and $c_{1}=(\text{number of 1's in }\beta)-1$, $c_{k}=(\text{number of k's in }\beta)$ for $k>1$.
        $\big[t^{2g}\big]$ means taking the coefficient of $t^{2g}$.
        \end{theorem}
        Here the Double Hurwitz number is counted with multiplicity, hence in our notation it will read as follows
        \begin{equation}\label{dbl_hurwitz_ourcase}
            \frac{H^{\bullet}_{\chi_{\infty}}((d),\nu)}{(-\chi_{\infty}+1+l(\nu))!}=
            \frac{d^{-\chi_{\infty}+l(\nu)}}{\vert\text{Aut }\nu\vert}
            \big[t^{2g}\big]\prod_{k\geq 1}\Big(\frac{\rm{sinh}\it(kt/2)}{kt/2}\Big)^{c_{k}}
        \end{equation}
        And in this case, (\ref{formula_recursion}) can be written as
        \begin{equation}\label{formula_simple_recursion}
            \sum_{\vert\nu\vert=d}\sum_{\chi_{0},\chi_{\infty}}\mathcal{D}^{\bullet}_{\chi_{0},\nu,e}
            z_{\nu}(-1)^{-\chi_{\infty}+1+l(\nu)}\frac{d^{-\chi_{\infty}+l(\nu)}}{\vert\text{Aut }\nu\vert}
            \big[t^{2g}\big]\prod_{k\geq 1}\Big(\frac{\rm{sinh}\it(kt/2)}{kt/2}\Big)^{c_{k}}=0
        \end{equation}
        Now fix $\nu$ and consider the case where there are $m$ vertices in $V(\Gamma)^{(0)}$. Then we have splitting of
        $\chi_{0}$, $\nu$, and $e$ into $\{g_{1},\cdots,g_{m}\}$, $\{\nu(v_{1}),\cdots,\nu(v_{m})\}$,
        and $\{e(v_{1}),\cdots,e(v_{m})\}$ such that $e(v_{i})$'s are allowed to be empty and
        $$\sum_{i=1}^{m}(2-2g_{i})=\chi_{0},\qquad \bigcup_{i=1,\cdots,m}\nu(v_{i})=\nu,\qquad\bigcup_{i=1,\cdots,m}e(v_{i})=e$$
        Each vertex will correspond to a certain Hodge integral on $\overline{\mathcal{M}}_{g(v),l(\nu(v))+l(e(v))}$ with dimension
        $3g(v)-3+l(\nu(v))+l(e(v))$. There are conditions on $m$, $\chi,\chi_{0},\chi_{\infty},l(e(v))$, and $l(\nu(v))$:
        let $\overline{g}(v)$ denote $\sum_{w\neq v}g(w)$,
        \begin{align*}
            &m\leq l(\nu),\quad l(\nu(v))\leq l(\nu)-m+1,
            &&\chi=\chi_{0}+\chi_{\infty}-2l(\nu),\quad l(e(v))\leq l(e)\\
            & \chi_{\infty}\leq 2\,\text{min }\{\,\,l((d)),l(\nu)\,\,\}=2, &&\chi_{0}=\sum_{i=1}^{m}(2-2g(v_{i}))=2m-2g(v)-2\overline{g}(v)
        \end{align*}
        From these conditions, we can deduce that
        \begin{align*}
            3g(v) &=\quad 3m-3\overline{g}(v)-\frac{3}{2}\chi_{0}
                =3m-3\overline{g}(v)-\frac{3}{2}\chi+\frac{3}{2}\chi_{\infty}-3l(\nu)\\
                &\leq -\frac{3}{2}\chi+3m-\frac{3}{2}\big(2l(\nu)-\chi_{\infty}\big)
        \end{align*}
        where equality holds if and only if $\overline{g}(v)=\sum_{w\neq v}g(w)=0$.
        Now we can find the upper bound for the dimension of Hodge integral as follows.\\
        \begin{align*}
            3g(v)-3 &+l(\nu(v))+l(e(v))
            \leq\quad-\frac{3}{2}\chi+3m-3-\frac{3}{2}\big(2l(\nu)-\chi_{\infty}\big)+l(\nu(v))+l(e) \\
            \leq&\quad\Big[3g-3+l(e)+1\Big]+\Big[3m-3-3l(\nu)+\frac{3}{2}\chi_{\infty}+l(\nu)-m\Big]\\
            \leq&\quad\Big[3g-3+l(e)+1\Big]+\Big[ 2\big(m-l(\nu)\big)+\frac{3}{2}\big(\chi_{\infty}-2\big)\Big]\\
            \leq&\quad3g-2+l(e)
        \end{align*}
        and the equality holds if and only if
        $$m=l(\nu),\quad \chi_{\infty}=2,\quad e(w)=\emptyset\text{ for all }w\neq v,\quad g(w)=0\text{ for all }w\neq v$$
        ,i.e. when each part of $\nu$ is splitted into separate vertices, and all the marked points other than the ramification divisor
        as well as all genuses are concentrated on one vertex on the $0$-th side.
        Now we can compute any Hodge integral with at-most one $\lambda$-class as follows:
        Say we want to compute Hodge integrals of the form
        $$\int_{\overline{M}_{g,n+1}}\psi_{0}^{k_{0}}\cdots\psi_{n}^{k_{n}}\lambda_{j}$$
        where $j+\sum_{i=0}^{n}k_{i} = 3g-2+n$. Assume $0\leq k_{0}\leq\cdots\leq k_{n}$ and let
        $e=(k_{1},\cdots,k_{n})$ and $\chi=2-2g$.
        Then for any positive integer $d$ such that $d>\chi+\vert e\vert-1$,
        the recursion formula (\ref{formula_simple_recursion}) 
        expresses the top-dimensional Hodge integrals in terms of
        lower-dimensional Hodge integrals as follows:
        \begin{align*}
            \sum_{\vert\nu\vert=d}\Big[\Big(\prod_{i=1}^{l(\nu)}&\frac{\nu_{i}^{\nu_{i}-1}}{\nu_{i}!}\Big)
            \frac{(-1)^{l(\nu)-1}d^{l(\nu)-2}}{n!}\sum_{i=1}^{l(\nu)}\frac{\nu_{i}^{2}}
            {\vert\text{Aut }\hat{\nu_{i}}\vert}
            \int_{\overline{\mathcal{M}}_{g,n+1}}
            \frac{\Lambda_{g}^{\vee}(1)\prod_{j=1}^{n}(1-\psi_{j})^{k_{j}}}{1-\nu_{i}\psi_{0}}\Big]\\
            =&\text{terms consisting of lower dimensional Hodge integrals only...}
        \end{align*}
        where $\text{Aut }\hat{\nu_{i}}$ is the automorphism group of the partition
        $\hat{\nu_{i}}=(\nu_{1},\cdots,\hat{\nu_{i}},\cdots,\nu_{l(\nu)})$.
        In this expression, the Hodge integral term expands to:
        \begin{align*}
            &\int_{\overline{\mathcal{M}}_{g,n+1}}
                \frac{\Lambda_{g}^{\vee}(1)\prod_{j=1}^{n}(1-\psi_{j})^{k_{j}}}{1-\nu_{i}\psi_{0}}\\
            &=\int_{\overline{\mathcal{M}}_{g,n+1}}
                \big(1-\lambda_{1}+\lambda_{2}+\cdots+(-1)^{g}\lambda_{g}\big)
                \big(\sum_{a_{0}=0}^{\infty}\nu_{i}^{a_{0}}\psi_{0}^{a_{0}}\big)
                \prod_{j=1}^{n}\Big[\sum_{a_{j}=0}^{k_{j}}(-1)^{a_{j}}{k_{j}\choose a_{j}}\psi_{j}^{a_{j}}\Big] \\
            &=\sum_{k+\sum a_{j}=3g-2+n}(-1)^{3g-2+n}\Big[\nu_{i}^{a_{0}}\prod_{j=1}^{n}
                {k_{j}\choose a_{j}}\Big] \int_{\overline{\mathcal{M}}_{g,n+1}}
                \psi_{0}^{a_{0}}\times\cdots\times\psi_{n}^{a_{n}}\lambda_{k}
        \end{align*}
        Hence the previous expression can be written as
        \begin{equation}\label{linear_relation}
            \sum_{k+\sum a_{j}=3g-2+n}C_{d}(k,(a_{j}))\int_{\overline{\mathcal{M}}_{g,n+1}}
            \psi_{0}^{a_{0}}\times\cdots\times\psi_{n}^{a_{n}}\lambda_{k}
            =\text{lower dimensional terms...}
        \end{equation}
        where $C_{d}(k,(a_{j}))$ are constants defined as follows:
        \begin{equation*}
            C_{d}\big(k,(a_{j})\big)=
            \sum_{\vert\nu\vert=d}\Big[\Big(\prod_{i=1}^{l(\nu)}\frac{\nu_{i}^{\nu_{i}-1}}{\nu_{i}!}\Big)
            \frac{(-1)^{l(\nu)-1}d^{l(\nu)-2}}{n!}\sum_{i=1}^{l(\nu)}\frac{\nu_{i}^{2}}
            {\vert\text{Aut }\hat{\nu_{i}}\vert}
            \Big(\nu_{i}^{a_{0}}\prod_{j=1}^{n}{k_{j}\choose a_{j}}\Big)\Big]
        \end{equation*}
        Now we have infinitely many linear relations of finitely many Hodge integrals of
        fixed dimension $3g-2+n$ with at-most one $\lambda$-class since
        the equation (\ref{linear_relation}) holds for all positive integers $d$ such that
        $d>1-2g+\sum k_{j}$. Moreover the coefficients $C_{d}(k,(a_{j}))$ form Vandermonde-type
        matrices and it can be proved that one can always find a set of positive integers
        $\{d_{1},\cdots,d_{l}\}$ which will give linearly independent relations to solve for all the
        Hodge integrals of given dimension $3g-2+n$ with at-most one $\lambda$-class 
        in terms of the values of lower-dim'l
        Hodge integrals with at-most one $\lambda$-class.
        So we just proved the following theorem;
        \begin{theorem}
            Any given Hodge integral with one $\lambda$-class:
            \begin{equation}\label{eqn:one_lambda}
                \int_{\overline{\mathcal{M}}_{g,n}}\psi_{1}^{k_{1}}\cdots\psi_{n}^{k_{n}}\lambda_{j}
            \end{equation}
            where $k_{1},\cdots,k_{n}\in\mathbb{N}\cup\{0\}$, $j\in\{0,1,2,\cdots,g\}$,
            is explicitly expressed as a polynomial in terms of lower-dimensional Hodge integrals with one $\lambda$-class.
            Therefore it computes all Hodge integrals with one $\lambda$-class.
        \end{theorem}
%-------------------------------------------------------------------------------
    \vskip 10mm
    \section{Algorithm to compute Hodge Integrals with one $\lambda$-class}
        It is clear that Theorem 7.2 can be implemented. We can use the formula (\ref{dbl_hurwitz_ourcase})
        to compute Double Hurwitz numbers. And since the values of Hodge integrals will be too big
        to fit in the usual 4-byte integer data type, we will need a library for multi-precision
        computing. There are several free libraries on the web and GNU-MP is one of them which
        provides well-organized C++ class interfaces as well as documentations.\\

        For any given Hodge integral
        $$\int_{\overline{\mathcal{M}}_{g,n}}\psi_{1}^{k_{1}}\cdots\psi_{n}^{k_{n}}\lambda_{j}$$
        Let $e=(k_{1},\cdots,k_{n-1})$ and $\chi=2-2g$. Start with $d=\chi-1+\vert e\vert,\cdots$
        and find linear relations (\ref{formula_recursion}) as follows:
        Run over partitions $\nu$ of size $d$. For a fixed $\nu$, run over pairs $(\chi_{0},\chi_{\infty})$
        which satisfies $\chi=\chi_{0}+\chi_{\infty}-2\,l(\nu)$, $\chi_{\infty}\leq 2$, $\chi_{0},\chi_{\infty}\in 2\,\mathbb{Z}$.
        Now for a fixed pair $(\chi_{0},\chi_{\infty})$, we can compute Double Hurwitz Number
        $\mathcal{D}_{\chi_{\infty}}^{\bullet}(\,(d)\, ,\nu)$ as follows:
        When $c_{1}\geq 0$, i.e. when there are one or more $1$'s in $\nu$, we have
        \begin{equation*}
            \big[ t^{2g}\big]\prod_{k\geq 1}\Big(\frac{\text{sinh}(kt/2)}{kt/2}\Big)^{c_{k}}
            =\sum_{(b_{k})}\prod_{b_{k}\neq 0}\sum_{(a_{j}^{k})}\frac{k^{2b_{k}}}{2^{2b_{k}}\prod_{j}(2a_{j}^{k}+1)!}
        \end{equation*}
        where $b_{k}=0$ if $c_{k}=0$ and $\chi_{\infty}=2-2g_{\infty}$, $\sum_{k} b_{k}=g_{\infty}$,
        $\sum_{j=1,\cdots,c_{k}} a_{j}^{k}=b_{k}$ with $b_{k}, a_{j}^{k}\in \mathbb{N}\cup\{0\}$.
        When $c_{1}=-1$, i.e. when there is no $1$ in $\nu$, let $h$ be the minimum numbered part of $\nu$.
        Then we have $c_{h}\geq 1$ and
        \begin{align*}
            &\Big(\frac{\text{sinh}(t/2)}{t/2}\Big)^{-1}\Big(\frac{\text{sinh}(ht/2)}{ht/2}\Big)
            =\frac{1}{h}\Big( e^{(h-1)t/2}+e^{(h-3)t/2}+\cdots+e^{(-h+1)t/2}\Big)\\
            &=\begin{cases}
                \sum_{m=0}^{\infty}\Big[\frac{1}{h\,2^{2m-1}(2m)!}\Big(\sum_{l}l^{2m}\Big)\Big]t^{2m},
                &\text{when }h\text{ is even, }l=1,3,\cdots,h-1 \\
                \frac{1}{h}+\sum_{m=0}^{\infty}\Big[\frac{1}{h\,2^{2m-1}(2m)!}\Big(\sum_{l}l^{2m}\Big)\Big]t^{2m},
                &\text{when }h\text{ is odd, }l=2,4,\cdots,h-1 \\
            \end{cases}
        \end{align*}
        Hence we have
        \begin{align*}
            \prod_{k\geq 1}\Big( & \frac{\text{sinh}(kt/2)}{kt/2}\Big)^{c_{k}}
            =\Big(\frac{\text{sinh}(t/2)}{t/2}\Big)^{-1}\Big(\frac{\text{sinh}(ht/2)}{ht/2}\Big)
            \Big(\frac{\text{sinh}(ht/2)}{ht/2}\Big)^{c_{h}-1}
            \prod_{k>h}\Big(\frac{\text{sinh}(kt/2)}{kt/2}\Big)^{c_{k}}\\
            =&\Big(\frac{1}{h}\delta_{h,\text{odd}}+\sum_{m=0}^{\infty}\Big[\frac{1}{h\,2^{2m-1}(2m)!}\Big(\sum_{l}l^{2m}\Big)\Big]t^{2m}\Big)
            \Big(\frac{\text{sinh}(ht/2)}{ht/2}\Big)^{c_{h}-1}
            \prod_{k>h}\Big(\frac{\text{sinh}(kt/2)}{kt/2}\Big)^{c_{k}}
        \end{align*}
        and in this case, for $g>0$;
        \begin{align*}
            \big[ t^{2g}\big]\prod_{k\geq 1}\Big(\frac{\text{sinh}(kt/2)}{kt/2}\Big)^{c_{k}}
            =&\sum_{(b_{k})}\Big[\frac{1}{h\,2^{2b_{1}-1}(2b_{1})!}
            \Big(\sum_{l}l^{2b_{1}}\Big)\Big]\prod_{b_{k}\neq 0,\,k\geq h}\sum_{(a_{j}^{k})}
            \frac{k^{2b_{k}}}{2^{2b_{k}}\prod_{j}(2a_{j}^{k}+1)!}
        \end{align*}
        where $b_{h}=0$ if $c_{h}=1$ and $b_{k}=0$ if $c_{k}=0$ for $k\neq h$.
        Using these formulas and (\ref{dbl_hurwitz_ourcase}), we can compute Double Hurwitz numbers.
        In order to compute $\big[\lambda^{l(\nu)-\chi_{0}}\big]\mathcal{D}^{\bullet}_{\nu,e}(\lambda)$,
        first run over the number of vertices $m=1,2,\cdots,l(\nu)$. For each $m$, find all possible
        groupings of $\nu$, $e$, and all possible splittings of $\chi_{0}$. Note that $e$ admits
        groupings with empty components. Now find all triples $(\nu(v),e(v),g(v))$ according to the
        equivalence condition of vertices discussed in Section 3. Then the contribution of
        $\big[\lambda^{l(\nu)-\chi_{0}}\big]\mathcal{D}^{\bullet}_{\nu,e}(\lambda)$ will be the product
        of the combination factor $1/\prod m_{i}!$ and the expansions of $\mathcal{D}_{g(v),\nu(v),e(v)}$
        which can be obtained by
        \begin{align*}
            &\int_{\overline{\mathcal{M}}_{g,l(\nu)+l(e)}}
            \frac{\Lambda^{\vee}_{g}(1)\prod_{j=1}^{l(e)}\big(1-\psi_{j}\big)^{e_{j}}}
            {\prod_{i=1}^{l(\nu)}\big(1-\nu_{i}\psi_{i}\big)}\\
            &=\int_{\overline{\mathcal{M}}_{g,l(\nu)+l(e)}}
            \big[1-\lambda_{1}+\cdots+(-1)^{g}\lambda_{g}\big]
            \prod_{j=1}^{l(e)}\big[\sum_{(\tilde{l}_{j})}(-1)^{\tilde{l}_{j}}{e_{j}\choose\tilde{l}_{j}}
            \psi_{j}^{\tilde{l}_{j}}\big]
            \prod_{i=1}^{l(\nu)}\big[\sum_{l_{i}}\nu_{i}^{l_{i}}\psi_{i}^{l_{i}}\big] \\
            &=\sum_{k,(l_{i}),(\tilde{l}_{j})}(-1)^{k+\sum\tilde{l}_{j}}\big[\prod_{j=1}^{l(e)}
            {e_{j}\choose\tilde{l}_{j}}\big]\big[\prod_{i=1}^{l(\nu)}\nu_{i}^{l_{i}}\big]
            \int_{\overline{\mathcal{M}}_{g,l(\nu)+l(e)}}
            \lambda_{k}\prod_{i=1}^{l(\nu)}\psi_{i}^{l_{i}}\prod_{j=1}^{l(e)}\psi_{j}^{\tilde{l}_{j}}
        \end{align*}
        where $l_{i}\geq 0$, $0\leq\tilde{l}_{j}\leq e_{j}$, $0\leq k\leq g$, and
        $k+\sum_{i}l_{i}+\sum_{j}\tilde{l}_{j}=3g-3+l(\nu)+l(e)$.
        Some of them will have maximum dimension $3g-3+n$ for the situations described in Section 7.
        Those are treated as unknowns and all others are lower-dimensional Hodge integrals or initial
        values which are already computed. Summing over all pairs $(\chi_{0},\chi_{\infty})$ and $\nu$
        will give a linear relation between Hodge integrals of the dimension $3g-3+n$. Now we can follow
        same step as above for other values of $d$ and obtain more linear relations. Observe that the
        number of unknowns are independent of $d$ and actually bounded by the number of partitions of
        $3g-3+n$, and hence we will have a system of linear relations which can be solved by simple
        Gaussian Elimination method. Thus in each dimension, it amounts to solve a matrix equation
        of size $N\times N$ when $N$ is at-worst-case the number of partitions of dimension.
%-------------------------------------------------------------------------------
    \vskip 10mm
    \section{Examples}
        In this section, I will show how the algorithm developed in the previous section works and check
        the results with the previously known-methods.\\

        \begin{itemize}
            \item[\underline{Dimension 1} : ]
                There are 3 Hodge integrals with one $\lambda$-class of dimension 1;
                \begin{equation*}
                \int_{\overline{\mathcal{M}}_{0,4}}\psi^{1},\qquad\qquad
                \int_{\overline{\mathcal{M}}_{1,1}}\lambda_{1},\qquad\qquad
                \int_{\overline{\mathcal{M}}_{1,1}}\psi^{1}
                \end{equation*}
                And these integrals can be computed as follows:
        \begin{align*}
        d=3, g=0,& e=(0, 0, 0);
        \qquad6\int_{\overline{\mathcal{M}}_{0,4}}\psi^{1}-6=0\implies\int_{\overline{\mathcal{M}}_{0,4}}\psi^{1}=1\\
        d=1, g=1,& e=\emptyset,\quad\text{and }d=2, g=1,e=\emptyset;\\
        &-\int_{\overline{\mathcal{M}}_{1,1}}\lambda_{1}+\int_{\overline{\mathcal{M}}_{1,1}}\psi^{1}=0,
        \qquad-\int_{\overline{\mathcal{M}}_{1,1}}\lambda_{1}+3\int_{\overline{\mathcal{M}}_{1,1}}\psi^{1}-\frac{1}{12}=0\\
        \implies&\int_{\overline{\mathcal{M}}_{1,1}}\lambda_{1}=\int_{\overline{\mathcal{M}}_{1,1}}\psi^{1}=\frac{1}{24}
        \end{align*}
        This value of $\psi$ class integral recovers the exceptional case (2.46) in \cite{Wit}, and the value of
        $\lambda_{1}$ class integral matches with the $\lambda_{g}$-formula since $B_{2}=1/6$.\\

            \item[\underline{Dimension 2} : ]
                There are 5 Hodge integrals with one $\lambda$-class of dimension 2;
                \begin{equation*}
                \int_{\overline{\mathcal{M}}_{0,5}}\psi^{2},\qquad
                \int_{\overline{\mathcal{M}}_{0,5}}\psi^{1}\psi^{1},\qquad
                \int_{\overline{\mathcal{M}}_{1,2}}\psi^{1}\lambda_{1},\qquad
                \int_{\overline{\mathcal{M}}_{1,2}}\psi^{2},\qquad
                \int_{\overline{\mathcal{M}}_{1,2}}\psi^{1}\psi^{1}
                \end{equation*}
                And these integrals can be computed as follows:
        \begin{align*}
        d=3, g=0,& e=(0, 0, 0, 0);\\
        &25\int_{\overline{\mathcal{M}}_{0,5}}\psi^{2}-25=0\implies\int_{\overline{\mathcal{M}}_{0,5}}\psi^{2} = 1\\
        d=4, g=0,& e=(1, 0, 0, 0);\\
        &-10\int_{\overline{\mathcal{M}}_{0,5}}\psi^{1}\psi^{1}+20=0\implies\int_{\overline{\mathcal{M}}_{0,5}}\psi^{1}\psi^{1} = 2\\
        d=1, g=1,& e=(0),\quad\text{and }d=2, g=1, e=(0);\\
        &-\int_{\overline{\mathcal{M}}_{1,2}}\psi^{1}\lambda_{1}+\int_{\overline{\mathcal{M}}_{1,2}}\psi^{2}=0,
        \qquad-3\int_{\overline{\mathcal{M}}_{1,2}}\psi^{1}\lambda_{1}+7\int_{\overline{\mathcal{M}}_{1,2}}\psi^{2}-\frac{1}{6}=0\\
        \implies&\int_{\overline{\mathcal{M}}_{1,2}}\psi^{1}\lambda_{1} =\int_{\overline{\mathcal{M}}_{1,2}}\psi^{2} = \frac{1}{24}\\
        d=2, g=1,& e=(1);\\
        &-3\int_{\overline{\mathcal{M}}_{1,2}}\psi^{1}\psi^{1}+\frac{1}{8}=0\implies\int_{\overline{\mathcal{M}}_{1,2}}\psi^{1}\psi^{1} = \frac{1}{24}
        \end{align*}
        The first two values matches with the Witten's formula for the $g=0$ case which says that
        \begin{equation}\label{Wit:gzero}
        \int_{\overline{\mathcal{M}}_{0,n}}\psi^{k_{1}}\cdots\psi^{k_{n}} = {n-3\choose k_{1},\cdots,k_{n}}
        ,\qquad\text{when }k_{1}+\cdots+k_{n}=n-3
        \end{equation}
        and the remaining values match with the results in \cite{Zuber}, and $\lambda_{g}$-formula (\ref{eqn:lambda_g}). \\

            \item[\underline{Dimension 3} : ]
                There are 8 Hodge integrals with one $\lambda$-class of dimension 3;
                \begin{align*}
                &\int_{\overline{\mathcal{M}}_{0,6}}\psi^{3},\qquad
                \int_{\overline{\mathcal{M}}_{0,6}}\psi^{2}\psi^{1},\qquad
                \int_{\overline{\mathcal{M}}_{0,6}}\psi^{1}\psi^{1}\psi^{1},\qquad
                \int_{\overline{\mathcal{M}}_{1,3}}\psi^{2}\lambda_{1},\\
                &\int_{\overline{\mathcal{M}}_{1,3}}\psi^{3},\qquad
                \int_{\overline{\mathcal{M}}_{1,3}}\psi^{1}\psi^{1}\lambda_{1},\qquad
                \int_{\overline{\mathcal{M}}_{1,3}}\psi^{2}\psi^{1},\qquad
                \int_{\overline{\mathcal{M}}_{1,3}}\psi^{1}\psi^{1}\psi^{1}
                \end{align*}
                And these integrals can be computed as follows:
        \begin{align*}
        d=3, g=0,& e=(0, 0, 0, 0, 0);\\
        &90\int_{\overline{\mathcal{M}}_{0,6}}\psi^{3}-90=0\implies\int_{\overline{\mathcal{M}}_{0,6}}\psi^{3} = 1\\
        d=4, g=0,& e=(1, 0, 0, 0, 0);\\
        &-65\int_{\overline{\mathcal{M}}_{0,6}}\psi^{2}\psi^{1}+195=0\implies\int_{\overline{\mathcal{M}}_{0,6}}\psi^{2}\psi^{1} = 3\\
        d=5, g=0,& e=(1, 1, 0, 0, 0);\\
        &15\int_{\overline{\mathcal{M}}_{0,6}}\psi^{1}\psi^{1}\psi^{1}-90=0\implies\int_{\overline{\mathcal{M}}_{0,6}}\psi^{1}\psi^{1}\psi^{1} = 6\\
        d=1, g=1,& e=(0, 0),\quad\text{and }d=2, g=1, e=(0, 0);\\
        &-\int_{\overline{\mathcal{M}}_{1,3}}\psi^{2}\lambda_{1}+\int_{\overline{\mathcal{M}}_{1,3}}\psi^{3}=0,
        \qquad-7\int_{\overline{\mathcal{M}}_{1,3}}\psi^{2}\lambda_{1}+15\int_{\overline{\mathcal{M}}_{1,3}}\psi^{3}-\frac{1}{3}=0\\
        \implies&\int_{\overline{\mathcal{M}}_{1,3}}\psi^{2}\lambda_{1} =\int_{\overline{\mathcal{M}}_{1,3}}\psi^{3} = \frac{1}{24}\\
        d=2, g=1,& e=(1, 0),\quad\text{and }d=3, g=1, e=(1, 0);\\
        &3\int_{\overline{\mathcal{M}}_{1,3}}\psi^{1}\psi^{1}\lambda_{1}-7\int_{\overline{\mathcal{M}}_{1,3}}\psi^{2}\psi^{1}+\frac{1}{3}=0,\\
        &6\int_{\overline{\mathcal{M}}_{1,3}}\psi^{1}\psi^{1}\lambda_{1}-25\int_{\overline{\mathcal{M}}_{1,3}}\psi^{2}\psi^{1}+\frac{19}{12}=0\\
        \implies&\int_{\overline{\mathcal{M}}_{1,3}}\psi^{1}\psi^{1}\lambda_{1} = \int_{\overline{\mathcal{M}}_{1,3}}\psi^{2}\psi^{1} = \frac{1}{12}\\
        d=3, g=1,& e=(1, 1);\\
        &6\int_{\overline{\mathcal{M}}_{1,3}}\psi^{1}\psi^{1}\psi^{1}-\frac{1}{2}=0\implies\int_{\overline{\mathcal{M}}_{1,3}}\psi^{1}\psi^{1}\psi^{1} = \frac{1}{12}
        \end{align*}
        The first 3 values match with (\ref{Wit:gzero}), and the rest match with
        $\lambda_{g}$-formula (\ref{eqn:lambda_g}) and the results in \cite{Zuber}, p36.\\

            \item[\underline{Dimension 4} : ]
                There are 16 Hodge integrals with one $\lambda$-class of dimension 4;
                \begin{align*}
                &\int_{\overline{\mathcal{M}}_{0,7}}\psi^{4},\qquad
                \int_{\overline{\mathcal{M}}_{0,7}}\psi^{3}\psi^{1},\qquad
                \int_{\overline{\mathcal{M}}_{0,7}}\psi^{2}\psi^{2},\qquad
                \int_{\overline{\mathcal{M}}_{0,7}}\psi^{2}\psi^{1}\psi^{1},\\
                &\int_{\overline{\mathcal{M}}_{0,7}}\psi^{1}\psi^{1}\psi^{1}\psi^{1},\qquad
                \int_{\overline{\mathcal{M}}_{1,4}}\psi^{3}\lambda_{1},\qquad
                \int_{\overline{\mathcal{M}}_{1,4}}\psi^{4},\qquad
                \int_{\overline{\mathcal{M}}_{1,4}}\psi^{2}\psi^{1}\lambda_{1},\\
                &\int_{\overline{\mathcal{M}}_{1,4}}\psi^{3}\psi^{1},\qquad
                \int_{\overline{\mathcal{M}}_{1,4}}\psi^{1}\psi^{1}\psi^{1}\lambda_{1},\qquad
                \int_{\overline{\mathcal{M}}_{1,4}}\psi^{2}\psi^{1}\psi^{1},\qquad
                \int_{\overline{\mathcal{M}}_{1,4}}\psi^{2}\psi^{2},\\
                &\int_{\overline{\mathcal{M}}_{1,4}}\psi^{1}\psi^{1}\psi^{1}\psi^{1},\qquad
                \int_{\overline{\mathcal{M}}_{2,1}}\psi^{2}\lambda_{2},\qquad
                \int_{\overline{\mathcal{M}}_{2,1}}\psi^{3}\lambda_{1},\qquad
                \int_{\overline{\mathcal{M}}_{2,1}}\psi^{4}
                \end{align*}
                And these integrals can be computed as follows:
        \begin{align*}
        d=3, g=0,& e=(0, 0, 0, 0, 0, 0);\\
        &301\int_{\overline{\mathcal{M}}_{0,7}}\psi^{4}-301=0\implies\int_{\overline{\mathcal{M}}_{0,7}}\psi^{4} = 1\\
        d=4, g=0,& e=(1, 0, 0, 0, 0, 0);\\
        &-350\int_{\overline{\mathcal{M}}_{0,7}}\psi^{3}\psi^{1}+1400=0\implies
            \int_{\overline{\mathcal{M}}_{0,7}}\psi^{3}\psi^{1} = 4\\
        d=5, g=0,& e=(2, 0, 0, 0, 0, 0);\\
        &140\int_{\overline{\mathcal{M}}_{0,7}}\psi^{2}\psi^{2}-840=0\implies
            \int_{\overline{\mathcal{M}}_{0,7}}\psi^{2}\psi^{2} = 6\\
        d=5, g=0,& e=(1, 1, 0, 0, 0, 0);\\
        &140\int_{\overline{\mathcal{M}}_{0,7}}\psi^{2}\psi^{1}\psi^{1}-1680=0\implies
            \int_{\overline{\mathcal{M}}_{0,7}}\psi^{2}\psi^{1}\psi^{1} = 12\\
        d=6, g=0,& e=(1, 1, 1, 0, 0, 0);\\
        &-21\int_{\overline{\mathcal{M}}_{0,7}}\psi^{1}\psi^{1}\psi^{1}\psi^{1}+504=0\implies
            \int_{\overline{\mathcal{M}}_{0,7}}\psi^{1}\psi^{1}\psi^{1}\psi^{1} = 24
        \end{align*}
        These values match with (\ref{Wit:gzero}).\\

        \begin{align*}
        d=1, g=1,& e=(0, 0, 0),\quad\text{and }d=2, g=1, e=(0, 0, 0);\\
        &-\int_{\overline{\mathcal{M}}_{1,4}}\psi^{3}\lambda_{1}+\int_{\overline{\mathcal{M}}_{1,4}}\psi^{4}=0,\\
        &-15\int_{\overline{\mathcal{M}}_{1,4}}\psi^{3}\lambda_{1}+31\int_{\overline{\mathcal{M}}_{1,4}}\psi^{4}-\frac{2}{3}=0\\
        \implies&\int_{\overline{\mathcal{M}}_{1,4}}\psi^{3}\lambda_{1} =\int_{\overline{\mathcal{M}}_{1,4}}\psi^{4} = \frac{1}{24}\\
        d=2, g=1,& e=(1, 0, 0),\quad\text{and }d=3, g=1, e=(1, 0, 0);\\
        &7\int_{\overline{\mathcal{M}}_{1,4}}\psi^{2}\psi^{1}\lambda_{1}-15\int_{\overline{\mathcal{M}}_{1,4}}\psi^{3}\psi^{1}+1=0,\\
        &25\int_{\overline{\mathcal{M}}_{1,4}}\psi^{2}\psi^{1}\lambda_{1}-90\int_{\overline{\mathcal{M}}_{1,4}}\psi^{3}\psi^{1}+\frac{65}{8}=0\\
        \implies&\int_{\overline{\mathcal{M}}_{1,4}}\psi^{2}\psi^{1}\lambda_{1} =\int_{\overline{\mathcal{M}}_{1,4}}\psi^{3}\psi^{1} = \frac{1}{8}\\
        d=3, g=1,& e=(1, 1, 0),\quad\text{and }d=4, g=1, e=(1, 1, 0);\\
        &-6\int_{\overline{\mathcal{M}}_{1,4}}\psi^{1}\psi^{1}\psi^{1}\lambda_{1}+25\int_{\overline{\mathcal{M}}_{1,4}}\psi^{2}\psi^{1}\psi^{1}-\frac{19}{4}=0,\\
        &-10\int_{\overline{\mathcal{M}}_{1,4}}\psi^{1}\psi^{1}\psi^{1}\lambda_{1}+65\int_{\overline{\mathcal{M}}_{1,4}}\psi^{2}\psi^{1}\psi^{1}-\frac{55}{4}=0\\
        \implies&\int_{\overline{\mathcal{M}}_{1,4}}\psi^{1}\psi^{1}\psi^{1}\lambda_{1} =
            \int_{\overline{\mathcal{M}}_{1,4}}\psi^{2}\psi^{1}\psi^{1} = \frac{1}{4}\\
        d=3, g=1,& e=(2, 0, 0);\\
        &25\int_{\overline{\mathcal{M}}_{1,4}}\psi^{2}\psi^{2}-\frac{25}{6}=0\\
        \implies&\int_{\overline{\mathcal{M}}_{1,4}}\psi^{2}\psi^{2} = \frac{1}{6}\\
        d=4, g=1,& e=(1, 1, 1);\\
        &-10\int_{\overline{\mathcal{M}}_{1,4}}\psi^{1}\psi^{1}\psi^{1}\psi^{1}+\frac{5}{2}=0\\
        \implies&\int_{\overline{\mathcal{M}}_{1,4}}\psi^{1}\psi^{1}\psi^{1}\psi^{1} = \frac{1}{4}
        \end{align*}
        These values match with \cite{Zuber}, p36. and the $\lambda_{g}$-formula (\ref{eqn:lambda_g}).\\

        \begin{align*}
        d=1, g=2,& e=\emptyset,\quad d=2, g=2, e=\emptyset,\quad\text{and }d=3, g=2, e=\emptyset;\\
        &\int_{\overline{\mathcal{M}}_{2,1}}\psi^{2}\lambda_{2}-\int_{\overline{\mathcal{M}}_{2,1}}\psi^{3}\lambda_{1}+\int_{\overline{\mathcal{M}}_{2,1}}\psi^{4}=0,\\
        &7\int_{\overline{\mathcal{M}}_{2,1}}\psi^{2}\lambda_{2}-15\int_{\overline{\mathcal{M}}_{2,1}}\psi^{3}\lambda_{1}+31\int_{\overline{\mathcal{M}}_{2,1}}\psi^{4}-\frac{1}{240}=0,\\
        &25\int_{\overline{\mathcal{M}}_{2,1}}\psi^{2}\lambda_{2}-90\int_{\overline{\mathcal{M}}_{2,1}}\psi^{3}\lambda_{1}+301\int_{\overline{\mathcal{M}}_{2,1}}\psi^{4}-\frac{5}{48}=0\\
        \implies&\int_{\overline{\mathcal{M}}_{2,1}}\psi^{2}\lambda_{2} = \frac{7}{5760},
        \quad\int_{\overline{\mathcal{M}}_{2,1}}\psi^{3}\lambda_{1} = \frac{1}{480},
        \quad\int_{\overline{\mathcal{M}}_{2,1}}\psi^{4} = \frac{1}{1152}
        \end{align*}
        The first value matches with $\lambda_{g}$-formula (\ref{eqn:lambda_g}), since we have
        \begin{align*}
            &\sum_{k=0}^{m}{m+1\choose k}B_{k}=0,\quad\text{for }m>0
            \implies B_{4}=-\frac{1}{30} \\
            &{2*2+1-3\choose 2}\frac{2^{2*2-1}-1}{2^{2*2-1}}\frac{\vert B_{2*2}\vert}{(2*2)!}=\frac{7}{5760}\\
        \end{align*}
        The second value matches with (\ref{eqn:g-1}), since we have
        \begin{align*}
            &b_{g}=\frac{2^{2g-1}-1}{2^{2g-1}}\frac{\vert B_{2g}\vert}{(2g)!},\quad\text{for }g>0
            \implies b_{1}=\frac{1}{24},\quad b_{2}=\frac{7}{5760}\\
            &b_{g}\sum_{i=1}^{2g-1}\frac{1}{i}-\frac{1}{2}
            \sum_{g_{1}+g_{2}=g}\frac{(2g_{1}-1)!(2g_{2}-1)!}{(2g-1)!}b_{g_{1}}b_{g_{2}}
            =\frac{7}{5760}\Big(1+\frac{1}{2}+\frac{1}{3}\Big)-\frac{1}{2}\frac{1}{3!}\Big(\frac{1}{24}\Big)^{2}
            =\frac{1}{480}
        \end{align*}
        And the last value matches with the result in \cite{Zuber}, p36.
        \end{itemize}


\begin{thebibliography}{}

\bibitem{Gou-Jac-Vai} I.P.~Goulden, D.M.~Jackson, A.~Vainshtein,
{\em The number of ramified coverings of the sphere by the torus and surfaces of higher genera},
Ann. of Comb. {\bf 4} (2000), 27-46.

\bibitem{Gou-Jac-Vak} I.P.~Goulden, D.M.~Jackson, R.~Vakil,
{\em Towards the Geometry of Double Hurwitz Numbers},
preprint, math.AG/0309440

\bibitem{Getzler} E.~Getzler,
{\em Topological recursion relations in genus 2},
in ``Integrable systems and algebraic geometry''(Kobe/Kyoto, 1997),
World Science Publishing, River Edge, NJ. 1998, pp 73-106

\bibitem{Faber} C.~Faber,
{\em Algorithms for computing intersection numbers on moduli spaces of curves, with an application to the class of the locus of Jacobians},
in New Trends in Algebraic Geometry (K.Hulek, F.Catanese, C.Peters and M.Reid, eds.), 93-109,
Cambridge University Press, 1999

\bibitem{Fab-Pan} C.~Faber, R.~Pandharipande,
{\em Hodge integrals, partition matrices, and the $\lambda_{g}$ conjecture},
Ann. of Math. (2) {\bf 157} (2003), no. 1, 97--124

\bibitem{Gra-Pan} T.~Graber, R.~Pandharipande,
{\em Localization of virtual classes},
Invent. Math. {\bf 135} (1999), no. 2, 487--518.

\bibitem{Gra-Vak2} T.~Graber, R.~Vakil,
{\em Relative virtual localization and vanishing of tautological classes on moduli spaces of curves},
preprint, math.AG/0309227.

\bibitem{Ion-Par1} E.-N. Ionel, T. Parker, 
{\em Relative Gromov-Witten invariants},
Ann. of Math. (2) {\bf 157} (2003), no. 1, 45--96. 

\bibitem{Ion-Par2} E.-N.~Ionel, T.~Parker,
{\em The symplectic sum formula for Gromov-Witten invariants},
preprint, math.SG/0010207.

\bibitem{Kon} M.~Kontsevich,
{\em Intersection theory on the moduli space of curves and the matrix Airy function},
Comm. Math. Phys. {\bf 147} (1992), no. 1, 1--23.

\bibitem{Li1} J.~Li,
{\em Stable Morphisms to singular schemes and relative stable morphisms},
J. Diff. Geom. {\bf 57} (2001), 509-578.

\bibitem{Li2} J.~Li,
{\em Relative Gromov-Witten invariants and a degeneration formula of Gromov-Witten invariants},
J. Diff. Geom. {\bf 60} (2002), 199-293.

\bibitem{Li-Zha-Zhe} A.M.~Li, G.~Zhao, Q.~Zheng,
{\em The number of ramified coverings of a Riemann surface by Riemann surface},
Comm. Math. Phys. {\bf 213} (2000), no. 3, 685--696.

\bibitem{LLZ1} C.-C.~Liu, K.~Liu, J.~Zhou,
{\em A proof of a conjecture of Mari\~no-Vafa on Hodge Integrals},
J. Differential Geom.  {\bf 65} (2003),  no. 2, 289--340. 

\bibitem{LLZ2} C.-C,~Liu, K.~Liu, J.~Zhou,
{\em A formula of two-partition Hodge integrals}, preprint, math.AG/0310272.

\bibitem{LLZ3} C.-C,~Liu, K.~Liu, J.~Zhou,
{\em Mari\~{n}o-Vafa formula and Hodge Integral Identities}, preprint, math.AG/0308015

\bibitem{Mac} I.G.~MacDonald,
{\em Symmetric functions and Hall polynomials},
2nd edition. Claredon Press, 1995.

\bibitem{Mar-Vaf} M.~Mari\~{n}o, C.~Vafa,
{\em Framed knots at large $N$},
Orbifolds in mathematics and physics (Madison, WI, 2001), 185--204,
Contemp. Math., 310, Amer. Math. Soc., Providence, RI, 2002.

\bibitem{Oko-Pan} A.~Okounkov, R.~Pandharipande,
{\em Hodge integrals and invariants of the unknot},
Geom. Topol. {\bf 8}  (2004), 675--699.

\bibitem{Wit1} E.~Witten,
{\em Quantum field theory and the Jones polynomial},
Commun. Math. Phys. {\bf 121} (1989) 351-399.

\bibitem{Wit} E.~Witten,
{\em Two-dimensional gravity and intersection theory on moduli space},
Surveys in differential geometry (Cambridge, MA, 1990),
243--310, Lehigh Univ., Bethlehem, PA, 1991.

\bibitem{Zho1} J.~Zhou,
{\em Hodge integrals, Hurwitz numbers, and symmetric groups}, preprint, math.AG/0308024.

\bibitem{Zho2} J.~Zhou,
{\em A conjecture on Hodge integrals}, preprint, math.AG/0310282.

\bibitem{Zho3} J.~Zhou,
{\em Localizations on moduli spaces and free field realizations of Feynman rules}, preprint, math.AG/0310283.

\bibitem{Zho4} J.~Zhou,
{\em Hodge Integrals and Integrable Hierarchies}, preprint, math.AG/0310408.

\bibitem{Mulase1} M.~Mulase,
{\em Complete Integrability of the Kadomtsev-Petviashvili Equation},
Adv. Math. \bf{54 }\rm (1984), 57-66.

\bibitem{Mulase2} M.~Mulase,
{\em Cohomological Structure in Soliton Equations and Jacobian Varieties},
J. Diff. Geom. \bf{19 }\rm (1984), 403-430.

\bibitem{Fren-Wang} I.B.~Frenkel, W.~Wang,
{\em Virasoro algebra and wreath product convolution},
J. Alg. \bf{242 }\rm (2001), 656-671

\bibitem{Kyoto} T.~Miwa, M.~Jimbo, E.~Date,
{\em Solitons. Idifferential equations, symmetries and infinite-dimensional algebras},
Cambride Tracs in Mathematics, 135, Cambridge University Press, 2000.

\bibitem{Mirz} M.~Mirzakhani,
{\em Simple geodesics and Weil-Petersson volumes of moduli spaces of bordered Riemann surfaces},
preprint, 2003.

\bibitem{Zuber} C.~Itzykson, J.-B.~Zuber,
{\em Combinatorics of the Modular Group II: The Kontsevich integrals},
preprint, hep-th/9201001

\end{thebibliography}
\end{document}